\documentclass[graybox]{svmult}
\usepackage[T1]{fontenc}
\usepackage[utf8]{inputenc}
\usepackage{type1cm}        
\usepackage{makeidx}         
\usepackage{graphicx}        
\usepackage{multicol}        
\usepackage[bottom]{footmisc}
\usepackage{newtxtext}        
\usepackage[varvw]{newtxmath}       
\usepackage[utf8]{inputenc}
\usepackage[english]{babel}
\usepackage{amsmath,amsthm,amsfonts}
\usepackage{multirow}
\usepackage{tikz}
\usepackage{verbatim}
\usepackage{todonotes}
\usepackage{footmisc}
\usepackage{tocloft} 
\usepackage{isotope}
\usepackage{braket}
\usepackage{siunitx}
\usepackage{xcolor} 
\DeclareSIUnit\torr{Torr} 
\DeclareSIUnit\gauss{G}
\usepackage[numbers,sort]{natbib}
\usepackage{hyperref}
\hypersetup{
    colorlinks=true,
    linkcolor=blue,
    filecolor=blue,      
    urlcolor=cyan,
    citecolor=blue,
    pdftitle={Bosonic quantum mixtures with competing interactions: quantum liquid droplets and supersolids},
    pdfpagemode=FullScreen,
    }

\makeindex             

\begin{document}

\title*{Bosonic quantum mixtures with competing interactions: quantum liquid droplets and supersolids}
\titlerunning{Bosonic quantum mixtures: quantum liquid droplets and supersolids} 

\author{Sarah Hirthe and Leticia Tarruell}

\institute{Sarah Hirthe \at ICFO - Institut de Ciencies Fotoniques, The Barcelona Institute of Science and Technology, 08860 Castelldefels (Barcelona), Spain, \email{sarah.hirthe@icfo.eu}
\and
Leticia Tarruell \at ICFO - Institut de Ciencies Fotoniques, The Barcelona Institute of Science and Technology, 08860 Castelldefels (Barcelona), Spain, and \at ICREA, Pg. Llu\'is Companys 23, 08010 Barcelona, Spain, \email{leticia.tarruell@icfo.eu}}

\maketitle

\abstract*{These lecture notes contain an introduction to quantum simulation of bosonic systems in the continuum, focusing on weakly interacting Bose-Bose mixtures with competing mean-field interactions. When the values of such interactions are fine-tuned to almost completely cancel the mean-field energy, quantum fluctuations become apparent and dominate the behavior of the system, stabilizing an ultradilute quantum liquid phase. An analogous situation appears in single-component dipolar quantum gases. We review the mechanism that gives rise to this exotic quantum liquid, which can form droplets that are self-bound in the absence of any external confinement, and discuss their properties and dynamics in both the mixture and the dipolar cases. In dipolar gases, arrays of dipolar droplets stabilized by quantum fluctuations can establish global phase coherence and form supersolids. In bosonic mixtures, supersolidity can emerge already at the mean-field level through spin-orbit coupling. We discuss the properties of such spin-orbit-coupled supersolids, comparing them to their dipolar counterparts. Specifically, we focus on their periodic density modulation, phase coherence, and peculiar excitation spectrum, which hosts both superfluid and crystal excitations. Finally, we conclude by discussing open research directions in the areas of quantum liquid droplets and spin-orbit-coupled supersolids, in particular at the interface of the two research topics.}

\abstract{These lecture notes contain an introduction to quantum simulation of bosonic systems in the continuum, focusing on weakly interacting Bose-Bose mixtures with competing mean-field interactions. When the values of such interactions are fine-tuned to almost completely cancel the mean-field energy, quantum fluctuations become apparent and dominate the behavior of the system, stabilizing an ultradilute quantum liquid phase. An analogous situation appears in single-component dipolar quantum gases. We review the mechanism that gives rise to this exotic quantum liquid, which can form droplets that are self-bound in the absence of any external confinement, and discuss their properties and dynamics in both the mixture and the dipolar cases. In dipolar gases, arrays of dipolar droplets stabilized by quantum fluctuations can establish global phase coherence and form supersolids. In bosonic mixtures, supersolidity can emerge already at the mean-field level through spin-orbit coupling. We discuss the properties of such spin-orbit-coupled supersolids, comparing them to their dipolar counterparts. Specifically, we focus on their periodic density modulation, phase coherence, and peculiar excitation spectrum, which hosts both superfluid and crystal excitations. Finally, we conclude by discussing open research directions in the areas of quantum liquid droplets and spin-orbit-coupled supersolids, in particular at the interface of the two research topics.}

\section{Introduction}
\label{sec:Intro}

These lecture notes focus on the quantum simulation of bosonic systems in the continuum, i.e. systems that are not subject to any externally imposed discrete structure. This absence of a lattice typically makes it more challenging to access regimes beyond mean-field  theory, which are the natural target of quantum simulation. In the continuum the kinetic energy is unbounded and, in general, interactions must therefore be made strong in order to reach strongly correlated regimes. 

Strongly interacting systems are readily realized in two-component Fermi gases, where large scattering lengths can be accessed while three-body losses are suppressed by Fermi statistics. The situation is more challenging for bosons, where three-body losses are intrinsically present. Nevertheless, a variety of approaches have been exploited to access and investigate strongly-interacting bosonic systems in the continuum~\cite{ChevyJPBAMOP2016}. Among the most widely used are experiments probing strongly interacting Bose gases on timescales shorter than the characteristic loss times~\cite{WildPRL2012, MakotynNatPhys2014, FletcherS2017, EigenNature2018}, as well as studies of impurities strongly coupled to a bosonic bath, known as the Bose polaron problem~\cite{HuPRL2016, JorgensenPRL2016, YanScience2020, SkouNP2021, EtrychPRX2025}.

In these lecture notes we instead focus on weakly interacting bosonic systems. In this regime, observing beyond-mean-field effects is intrinsically challenging, since they appear only as small corrections on top of a dominant mean-field background. Revealing them therefore requires high‑precision measurements, e.g., of the equation-of‑state of the system~\cite{NavonPRL2011}. An alternative strategy to reveal beyond-mean-field effects in weakly interacting systems, and to make them play a central role in the physics, was proposed in 2015 by D.~S.~Petrov~\cite{PetrovPRL2015}. The key idea is to consider systems with two interactions of different origins and opposite signs that can be controlled independently. By fine-tuning their relative strengths, the overall mean-field interaction can be made to vanish. In this situation quantum fluctuations remain small in absolute terms, but their relative importance is dramatically enhanced and they can even dominate the behavior of the system.

In his seminal proposal, Petrov considered a regime where the residual mean-field interaction is attractive. The leading beyond-mean-field correction to the energy, which arises from quantum fluctuations, is instead always repulsive and gives rise to an effective interaction term known as the Lee-Huang-Yang energy correction. Remarkably, these two contributions can balance each other and stabilize a self-bound state of matter that Petrov termed a quantum liquid droplet. Such a quantum liquid is highly unconventional. First, it is universal: its properties do not depend on the range of the microscopic interaction potential and therefore do not follow the standard Van der Waals paradigm of liquids. Second, it provides a striking manifestation of quantum fluctuations, as its very existence relies on them even though the system remains weakly interacting. Finally, its equilibrium properties are largely determined by quantum fluctuations and correlations~\cite{PetrovNatPhys2018}.

Shortly after this theoretical proposal, quantum liquid droplets were observed experimentally~\cite{KadauNature2016, Ferrier-BarbutPRL2016}. Surprisingly, the first realization did not occur in the Bose-Bose mixture configuration originally envisioned by Petrov, but instead in a single-component dipolar quantum gas. In this case two mean-field interaction terms, the magnetic dipole-dipole interaction and the contact interaction, can also be tuned to nearly cancel each other, thereby revealing the role of quantum fluctuations. Quantum liquid droplets were subsequently realized in the original mixture setting~\cite{CabreraScience2018, SemeghiniPRL2018}. More recently, related physics has also been explored in Bose-Einstein condensates of polar molecules interacting \emph{via} electric dipole-dipole interactions~\cite{ZhangArXiv2025}. Although these lecture notes focus primarily on Bose-Bose mixtures, the physics of dipolar gases is closely related. We will therefore discuss the two platforms and their most important experiments in parallel. We will not attempt to give an exhaustive overview of dipolar experiments, as an excellent review of this field already exists~\cite{BöttcherRPP2021}.

Competing interactions do not only stabilize quantum liquid phases; more generally, they provide a powerful mechanism for engineering new states of matter whose existence relies crucially on quantum fluctuations. Several examples have been realized or proposed in ultracold quantum gases. In Bose-Bose mixtures, a gas phase of unconventional equation of state is stabilized by quantum fluctuations beyond the mean-field collapse threshold~\cite{CabreraScience2018, SemeghiniPRL2018, SkovPRL2021}. In the presence of a coherent coupling between the two components, a gas with emerging three-body interactions is also expected~\cite{LavoinePRL2021}. In dipolar gases, the establishment of global phase coherence in arrays of quantum droplets stabilized by quantum fluctuations has led to the stabilization of an extremely exotic phase of matter, the supersolid~\cite{BöttcherRPP2021,chomazRoPP2022, RecatiNRP2023}. 

Supersolidity refers to the coexistence of crystalline order and superfluidity within a single system. A supersolid breaks both the global U(1) gauge symmetry and translational symmetry, combining global phase coherence and a spontaneously modulated density profile~\cite{BonisegniPRMP2012}. Remarkably, such behavior has been observed not only in dipolar gases \cite{TanziPRL2019, BöttcherPRX2019, ChomazPRX2019}, but also in two other quantum-gas platforms: condensates in optical cavities \cite{LeonardNature2017, LéonardS2017, SchusterPRL2020} and spin-orbit-coupled Bose mixtures~\cite{LiNature2017, PutraPRL2020, ChisholmScience2026}. 

In bosonic mixtures with contact interactions, which constitute the main focus of these lecture notes, the supersolid phase emerges from spin-orbit coupling. Experimentally, spin-orbit coupling is realized using two-photon Raman transitions~\cite{LinN2011}. This coupling modifies the single-particle dispersion relation of the atoms and enables the engineering of a band whose momentum states have different spin composition and interaction properties~\cite{GalitskiNature2013}. When the lowest Raman-coupled band develops two minima, the gas can be viewed as an effective mixture of condensates whose spin components are not orthogonal~\cite{HigbiePRA2004, LinN2011}. Matter-wave interference between these two components then produces a spatially modulated density that spontaneously breaks translational symmetry. At the same time, global phase coherence is preserved, so the system retains its superfluid character and realizes a supersolid phase~\cite{HoPRL2011, LiARCAM2015, MartoneEPL2023}.

The supersolid phase realized in spin‑orbit-coupled mixtures shares many qualitative features with its dipolar counterpart, but it has a fundamentally different microscopic origin. Moreover, while spin-orbit-coupled supersolids arise within a mean-field framework, the dipolar supersolid phase emerges in the beyond-mean-field regime from arrays of quantum droplets stabilized by quantum fluctuations where, for suitable parameters, particle exchange between the droplets can establish global phase coherence across the array. The experimental demonstration of supersolidity in dipolar droplet arrays has therefore represented a major success in the field of quantum gases with competing interactions. Both dipolar and spin-orbit-coupled systems have since then enabled extensive studies of supersolid physics, ranging from the initial observation of simultaneous phase coherence and crystalline order~\cite{LiNature2017, TanziPRL2019, BöttcherPRX2019, ChomazPRX2019, PutraPRL2020} to detailed investigations of their collective excitations~\cite{TanziN2019, GuoN2019, NatalePRL2019, TanziScience2021, HertkornPRX2021, NorciaPRL2022, BiagioniNature2024, CasottiN2024, ChisholmScience2026}. Several excellent reviews already exist on supersolidity in dipolar quantum gases~\cite{BöttcherRPP2021, chomazRoPP2022, RecatiNRP2023}. Here we instead review the progress on supersolidity in spin-orbit-coupled mixtures, relating it to the dipolar case and discussing possible future research directions.

These lecture notes are organized in two parts. The first part focuses on Bose gases with competing interactions and quantum liquid droplets, and the second one covers supersolidity. We begin the quantum liquid droplet section with a review of the mean-field and beyond-mean-field description of single-component Bose gases, which we then extend to the relevant scenario of bosonic quantum mixtures. Next, we introduce the physics of competing interactions and the mechanism that stabilizes quantum liquid droplets in weakly interacting Bose gases, starting with Bose-Bose mixtures and subsequently discussing the dipolar case. We also review experimental realizations of quantum liquid droplets in both platforms, along with other beyond-mean-field experiments in the gas phase. This part concludes by highlighting open research directions in the field of quantum liquid droplets and related topics. 

In the second part of the lecture notes, we turn to supersolidity. After a brief introduction to supersolidity in quantum gases, we review its realization in dipolar systems before examining spin-orbit-coupled Bose gases. We cover the fundamentals of spin-orbit coupling, the phase diagram of spin-orbit-coupled condensates, and the origin of their supersolid (stripe) phase. Special emphasis is placed on describing these systems in terms of an effective mixture, which provides a unifying perspective across the two parts of the lectures. We then present the experimental study of spin-orbit-coupled supersolids, comparing the results with those obtained in dipolar gases and discussing open research directions for future research. Finally, the lecture notes conclude by discussing potential future directions that further connect the physics of quantum liquid droplets and supersolidity.

\section{Quantum liquid droplets in attractive Bose-Bose mixtures} \label{sec:Droplets}
The first superfluid ever realized was liquid helium, which, as its name suggests, is a liquid. Ultracold atomic systems, in contrast, are almost always encountered in a gaseous form. At first sight, this difference might seem fundamental. However, an ultracold quantum gas can under very specific conditions also form a liquid phase.

Stabilizing a liquid requires a delicate balance between attractive and repulsive forces: attraction binds the particles together, while repulsion prevents the system from collapsing. In weakly interacting quantum gases, such a balance can arise when two interaction contributions of different physical origins and opposite signs are finely tuned so that the overall mean-field energy nearly vanishes. In this situation, a small residual attraction remains at the mean-field level. Quantum fluctuations provide an effective repulsive contribution beyond the mean-field description that can then stabilize the system, giving rise to a novel form of matter: an ultradilute quantum liquid that appears in the form of self-bound quantum droplets.

In this section, we present the mechanisms that lead to the formation of such ultradilute liquids. We begin by reviewing the standard mean-field and beyond-mean-field descriptions of Bose-Einstein condensates (BECs) and of Bose-Bose mixtures. We then introduce the mechanism responsible for stabilizing the liquid phase in a mixture of BECs. After discussing the main properties of quantum liquid droplets in this setting, we show how closely related physics also emerges in dipolar condensates. Finally, we review the key experiments performed in these systems and conclude the section by highlighting open questions and promising directions for future research on ultradilute quantum liquids.

\subsection{Mean-field and beyond mean-field description of BECs with contact interactions} \label{subsec:MF and BMF single component}
While Bose-Einstein condensation is a purely quantum statistical effect that does not require inter-particle interactions, their presence is required for superfluidity and strongly influences the properties of BEC. In conventional BECs, those interactions stem from collisions between the atoms and can be described by a model contact potential of the form $V(r) = g \delta(r)$, where $r$ is the inter-atomic separation distance, $\delta(r)$ the Dirac delta function, and $g = 4\pi h^2a/m$ the contact coupling constant. Here $\hbar$ is Planck's constant, $a$ is the scattering length and $m$ is the atomic mass. 

Within the mean-field approximation, which assumes that all bosons occupy the same quantum state, the BEC is characterized by a macroscopic wave function $\psi(r,t)$. For a three-dimensional ensemble of spin-polarized bosons of density $n=\left|\psi\right|^2$ and confined in an external potential $V_{\mathrm{trap}}(r)$, the condensate wavefunction is the solution of the Gross-Pitaevksii equation:
\begin{equation}
i\hbar\partial_t \psi= \left[-\frac{\hbar^2}{2m}\nabla^2+V_{\mathrm{trap}}+ g |\psi|^2\right]\psi.\label{eq:GPE}
\end{equation}

At the mean-field level, the energy per unit volumen (or energy density) of the gas in the homogeneous limit reads $\mathcal{E}(\mathbf{r})=g n(\mathbf{r})^2/2$. It is related to the overall energy of the system by $E=\int \mathrm{d}\mathbf{r} \cal{E}(\mathbf{r})$. The expression of the energy already reveals that, while for attractive interactions ($g < 0$) the system collapses on itself, for repulsive interactions ($g > 0$) the BEC exists in a gas phase and minimizes its equation of state at the lowest density.

The mean-field approximation is valid only at low temperature and in the dilute limit, where the condition on the so-called gas parameter $na^3\ll1$ holds. When the temperature or the interactions are increased, thermal fluctuations and quantum fluctuations of the low-energy Bogoliubov modes deplete the condensate and lead to a modification of its equation of state. At zero-temperature, the first beyond-mean field corrections to the condensate energy were calculated in the 50's by Lee, Huang and Yang \cite{LeePR1957}. Including them, the energy density of the gas becomes
\begin{equation}
\mathcal{E}=\frac{1}{2}g n^2\left(1+\frac{128}{15\sqrt{\pi}}\sqrt{n a^3}+...\right)\label{eq:SingleComponentLHY}.
\end{equation}
This Lee-Huang-Yang (LHY) correction stems from the zero-point energies of Bogoliubov quasiparticles and is therefore always repulsive. In 3D systems it has a stronger density dependence than the mean-field term, as the energy density scales as $n^{5/2}$ instead of $n^2$. Interestingly, the LHY contribution to the chemical potential term scales with the gas parameter in the same way as the quantum depletion of the condensate state. 

\begin{figure}[ht]
\centering
\includegraphics[width=0.65\textwidth]{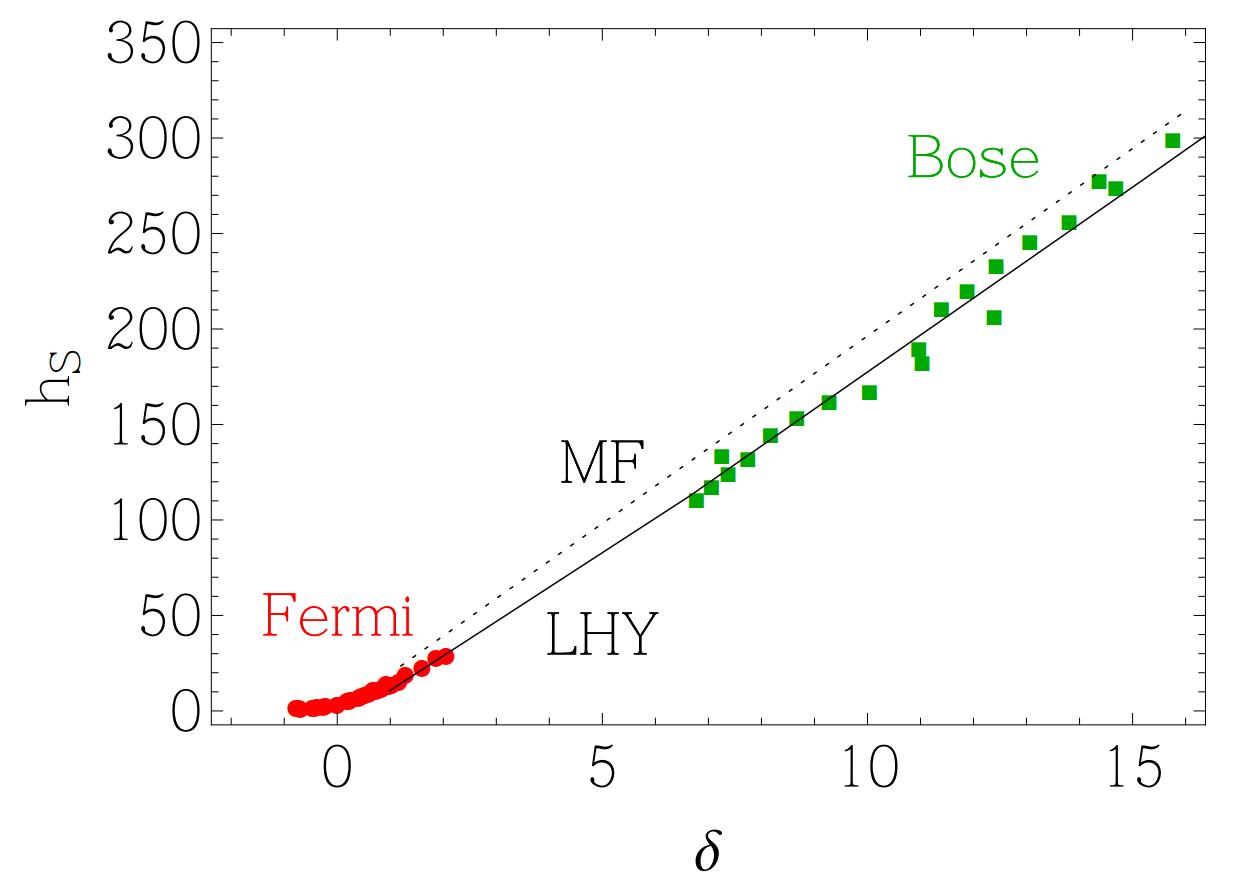}
\caption{\textbf{Experimental determination of the LHY correction.} Grand-canonical equation of state of a single-component Bose gas of $^7$Li atoms (green squares) and of a gas of weakly-bound bosonic molecules formed by two $^6$Li atoms (red circles). In the figure, $h_S=P/(2P_0)$ is the pressure of the gas normalized by that of an ideal Fermi gas $P_0$, and $\delta=\hbar/\sqrt{2m\tilde{\mu}}a$ is an interaction parameter. The latter is proportional to the scattering length $a$ and depends on the chemical potential $\tilde{\mu}$, from which the energy of the molecular bound state has been subtracted in the fermionic case. The grand-canonical equation of state $P(n)$ can be related to the more widely used canonical one $\mu(n)$ though thermodynamical relations. Figure extracted from \cite{NavonPhD2011}.}
\label{fig:EoSBFM} 
\end{figure}

Such LHY energy correction has been experimentally verified in several experiments, exploiting the ability to vary the scattering length of the gas to rather large values using Feshbach resonances. The most accurate measurements of the LHY term rely on the determination of the equation of state of a Bose gas through its density profile and yield its equation of state in the grand-canonical ensemble, that is, the pressure $P(n)$. As depicted in Fig. \ref{fig:EoSBFM}, such measurements have been carried out both in single component Bose gases \cite{NavonPRL2011}, and in Fermi gases where bosonic molecules are formed by binding two fermionic atoms \cite{NavonScience2010}. The two systems yield remarkable agreement with the LHY predictions, allowing to verify its validity in a broad range of interaction strengths. Alternatively, the LHY correction has also been investigated through spectroscopic measurements of strongly-interacting Bose gases, exploiting the fact that the response of the system to radio-frequency excitation in the regime of large detuning yields the 2-body contact $C$. This quantity is nothing else than the derivative of the equation of state of the gas with respect to $1/a$, and therefore yields an independent measurement of the LHY term \cite{WildPRL2012}.  

\subsection{Quantum mixtures with contact interactions: mean-field description and quantum fluctuations} \label{subsec:MF and BMF mixtures}
In single-component Bose gases, the LHY contribution to the equation of state is a correction to the mean-field result. Determining its value therefore requires high precision measurements as discussed above, and is challenging to achieve because the mean-field term provides a large background. In contrast, mixtures of Bose-Einstein condensates allow one to isolate the role of beyond mean-field terms and make them play a dominant role even in the weakly interacting regime. Such mixtures can be obtained either by using two different atomic species, which inevitably increases the experimental overhead, or more readily by exploiting two different internal states of the same atom. 
To understand how bosonic quantum mixtures make beyond mean-field effects more easily accessible, let us consider the interactions of such a system. If we denote the two components of the mixture as $\uparrow$ and $\downarrow$, three coupling constants characterize the system:  $g_{\uparrow\uparrow}$, $g_{\downarrow\downarrow}$, and $g_{\uparrow\downarrow}$. They are associated to three scattering lengths: $a_{\uparrow\uparrow}$ and $a_{\downarrow\downarrow}$ for the intraspin interactions and $a_{\uparrow\downarrow}$ for the interspin ones. 

\begin{figure}[b]
\centering
\includegraphics[scale=1]{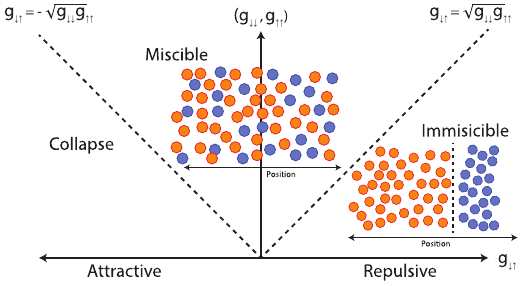}
\caption{\textbf{Mean-field phase diagram of a quantum mixture.} The phase diagram of a bosonic mixture where the two components are denoted as $\uparrow$ and $\downarrow$ is controlled by the three coupling constants $g_{\uparrow \uparrow}$, $g_{\downarrow \downarrow}$, and $g_{\uparrow \downarrow}$. For interspin interactions $-\sqrt{g_{\uparrow\uparrow} g_{\downarrow\downarrow}}<g_{\uparrow\downarrow}<\sqrt{g_{\uparrow\uparrow} g_{\downarrow\downarrow}}$, the system remains miscible. The condition $\left|g_{\uparrow\downarrow}\right|=\sqrt{g_{\uparrow\uparrow}g_{\downarrow\downarrow}}$, indicated in the figure by the diagonal dashed lines, marks the instability of the gas towards collapse (for the attractive $g_{\uparrow\downarrow}<0$ case) or phase separation (for the repulsive $g_{\uparrow\downarrow}>0$ case). Figure extracted from \cite{CabreraPhD2018}.}
\label{fig:MFPhaseDiagram} 
\end{figure}

At the mean-field level, the total interaction energy density of the mixture is 
\begin{equation}
\mathcal{E}=\frac{1}{2}\left(g_{\uparrow\uparrow} n^2_{\uparrow}+g_{\downarrow\downarrow} n^2_{\downarrow}\right)+g_{\uparrow\downarrow}n_{\uparrow}n_{\downarrow},\label{eq:EMFmixture}
\end{equation}
where we have denoted the densities of the two components $n_{\uparrow\uparrow}$ and $n_{\downarrow\downarrow}$. If $g_{\uparrow\uparrow}>0$ and  $g_{\downarrow\downarrow}>0$, each component is stable against collapse. Depending on the sign of $g_{\uparrow\downarrow}$, two situations can then occur. They are depicted in Fig. \ref{fig:MFPhaseDiagram}:
\begin{itemize}
    \item For $g_{\uparrow\downarrow}>0$ the two components repel each other. The lowest-energy excitations of the system are spin modes, in which the two spins are excited out of phase. The system remains miscible provided $g_{\uparrow\downarrow}^2< g^2$, where we have defined $g=\sqrt{g_{\uparrow\uparrow} g_{\downarrow\downarrow}}$ \cite{Pethick2008}. Beyond this limit, phase separation occurs. We will return to this scenario in Sec. \ref{sec:Supersolid}, in the context of spin-orbit-coupled supersolids.
    \item For $g_{\uparrow\downarrow}<0$ the two components attract each other, and interspin and intraspin interactions compete. The lowest-energy excitations of the system are density modes, in which the two components are excited in phase. The mixture can be conveniently described using the parameter $\delta g=g_{\uparrow\downarrow}+g$, which gives the strength of the remaining mean-field interactions. For $\delta g>0$, the system remains stable, albeit with very small overall mean-field interaction energy. For $\delta g<0$, a density instability that results in collapse is expected at the mean-field level. 
\end{itemize}

The scenario described above can be more deeply understood through a Bogoliubov analysis, considering the excitation spectrum of the system in the homogeneous limit. It is characterized by two Bogoliubov branches which, assuming that the two atoms have the same mass\footnote{All our subsequent discussion assumes this equal-mass condition, which simplifies the equations without modifying the essential physics. The unequal-mass results are given in \cite{PetrovPRL2015}.} ($m=m_{\uparrow}=m_{\downarrow}$), read
\begin{equation}
E(k)_{\pm}=\sqrt{\frac{\hbar^2 k^2}{2 m}\left(\frac{\hbar^2 k^2}{2 m}+2mc_{\pm}^2\right)},
\end{equation}
with the two speeds of sound
\begin{equation}
c^2_{\pm}=\frac{(g_{\uparrow\uparrow}n_{\uparrow}+g_{\downarrow\downarrow}n_{\downarrow})\pm\sqrt{(g_{\uparrow\uparrow}n_{\uparrow}-g_{\downarrow\downarrow}n_{\downarrow})^2+4g^2_{\uparrow\downarrow}n_{\uparrow}n_{\downarrow}}}{2 m}.
\end{equation}
There are therefore two characteristic healing lengths for the mixture $\xi_{\pm}=\hbar/(\sqrt{2}m c_{\pm})$, which represent the smallest length scales below which each branch cannot be excited. 

\begin{figure}[b]
\centering
\includegraphics[scale=1]{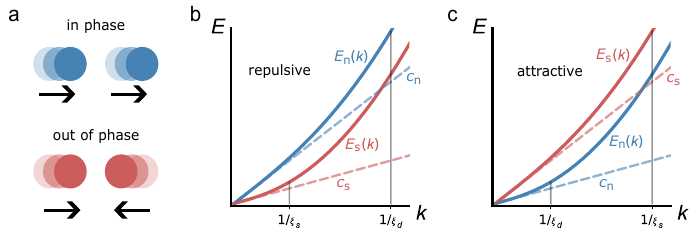}
\caption{\textbf{Bogoliubov excitation spectrum of a quantum mixture.}  (a)The excitation spectrum of a two-component Bose gas with contact interactions features two distinct Bogoliubov branches: density excitations, characterized by in-phase motion of the components, and spin excitations, where the components move out of phase. Each branch has a sound velocity and a healing length associated to it (dashed lines and vertical gray lines in (b) and (c), respectively). Here we depict the simpler case of symmetric intraspin interactions $g_{\uparrow\uparrow}= g_{\downarrow\downarrow}=\bar{g}$. (b) When interspin interactions are repulsive $g_{\uparrow\downarrow}>0$, the Bogoliubov branch of lowest energy corresponds to spin excitations. For $c_s<0$ the energy branch becomes imaginary and the system becomes unstable towards phase separation. (c) The situation is reversed in the case of attractive interspin interactions ($g_{\uparrow\downarrow}<0$), for which the lowest energy branch is the density one. In this case the instability, marked by $c_n<0$, is towards collapse.}
\label{fig:BogoliubovMixture} 
\end{figure}

If intraspin interactions are symmetric  $g_{\uparrow\uparrow}=g_{\downarrow\downarrow}=\bar{g}$ (the mixture has {\it balanced} interactions), the densities of the two spin components become equal $n_{\uparrow}=n_{\downarrow}=n/2$. The two Bogoliubov branches acquire a pure density or spin character, see Fig. \ref{fig:BogoliubovMixture}, and the speeds of sound and healing lengths take the simplified form 
\begin{eqnarray}
c_n=\sqrt{\frac{(\bar{g}+g_{\uparrow\downarrow})n}{2m}}\quad &\text{and}& \quad c_s=\sqrt{\frac{(\bar{g}-g_{\uparrow\downarrow})n}{2m}}\\
\xi_{n}=\frac{\hbar}{\sqrt{m n (\bar{g} +g_{\uparrow\downarrow})}}\quad &\text{and}& \quad\xi_{s}=\frac{\hbar}{\sqrt{m n (\bar{g} -g_{\uparrow\downarrow})}},
\end{eqnarray}
where we have used the notation $+\rightarrow n$ and $-\rightarrow s$. Instead, for asymmetric (imbalanced) interactions $g_{\uparrow\uparrow}\neq g_{\downarrow\downarrow}$ the two channels hybridize and have a mixed density-spin character.

In this mean-field picture, instabilities appear when the energy of the lowest Bogoliubov branch acquires an imaginary component. In the repulsive case, this is a spin branch and the system becomes unstable when the spin speed of sound $c_s$ becomes imaginary. Fluctuations leading to the separation of the two components are amplified and the system phase-separates. In contrast, in the attractive case it is the density branch that softens and the density speed of sound $c_n$ that becomes imaginary. The system becomes then unstable towards collapse, as fluctuations leading to the clumping of atoms in the same location of space are amplified. 

This picture is however considerably modified by beyond-mean-field effects in the attractive case. There, quantum fluctuations provide an additional repulsive contribution that is able to stabilize the system beyond the mean-field collapse threshold. To understand the situation, let’s 
have a closer look at the mean-field energy density of the mixture in the case of attractive interactions Eq. \eqref{eq:EMFmixture}, and rewrite it in terms of $g=\sqrt{g_{\uparrow\uparrow}g_{\downarrow\downarrow}}$ and $\delta g=g_{\uparrow\downarrow}+g$

\begin{equation}
\mathcal{E}=\frac{1}{2}\left(\sqrt{g_{\uparrow\uparrow}} n_{\uparrow}-\sqrt{g_{\downarrow\downarrow}} n_{\downarrow}\right)^2+\delta g n_{\uparrow}n_{\downarrow}.\label{eq:EMFmixturegdeltag}
\end{equation}

We are interested in the behavior of the gas in the vicinity of the collapse condition $\delta g\rightarrow 0$, and with $\delta g\ll g_{\uparrow\uparrow},g_{\downarrow\downarrow}$. There, the first term of Eq. \eqref{eq:EMFmixturegdeltag} is much larger than the second one.  The system will therefore minimize its energy by adjusting the densities of the two spin components such that their ratio satisfies
$n_{\uparrow}/n_{\downarrow}\simeq\sqrt{g_{\downarrow\downarrow}/g_{\uparrow\uparrow}}.$
This condition maximizes their spatial overlap. Assuming that the spin modes of the system are not excited, the mean-field energy density of the mixture then becomes

\begin{equation}
\mathcal{E}\simeq\frac{g\delta g}{\left(\sqrt{g_{\uparrow\uparrow}}+\sqrt{g_{\downarrow\downarrow}}\right)^2}n^2 \quad\mathrm{if}\quad n_{\uparrow}/n_{\downarrow}\simeq\sqrt{g_{\downarrow\downarrow}/g_{\uparrow\uparrow}},\label{eq:EMFsoft}
\end{equation}
where $n=n_{\uparrow}+n_{\downarrow}$. In this expression, we see that at the mean-field level the mixture can be effectively described as a single-component Bose gas of density $n$  and effective scattering length $2a\delta a /\left(\sqrt{a_{\uparrow\uparrow}} +\sqrt{a_{\downarrow\downarrow}}\right)^2$. However, the excitation spectrum of the mixture contains two Bogoliubov branches, as opposed to the excitation spectrum of a single component gas which has only one. The beyond-mean-field energy term, which is the zero-point energy corresponding to the Bogoliubov modes of the system, is thus very different. 

The LHY energy of a Bose-Bose mixture is computed from the energy of the two Bogoliubov branches through the momentum-space integral
\begin{eqnarray}
\mathcal{E}_{\mathrm{LHY}}&=&\int\frac{d^3k}{2 (2\pi)^3} [E_+(k)+E_-(k)-\frac{\hbar^2k^2}{m}-g_{\uparrow\uparrow}n_{\uparrow}-g_{\downarrow\downarrow}n_{\downarrow}+ \nonumber \\
&+& \frac{m}{\hbar^2k^2}\left(g_{\uparrow\uparrow}^2n_{\uparrow}^2+g_{\downarrow\downarrow}^2 n_{\downarrow}^2+2 g_{\uparrow\downarrow}^2n_{\uparrow}n_{\downarrow}\right)
].
\label{eq:MixtureLHY}
\end{eqnarray}

The main contribution to this integral comes from momenta of the order of the healing lengths. It can be conveniently rewritten as a function of the sound velocities, taking the explicit form \cite{LarsenAP1963}

\begin{equation}
\mathcal{E}_{\mathrm{LHY}}=\frac{8}{15\pi^2}\frac{m^4}{\hbar^3}\sum_{\pm}c^5_{\pm}=\frac{8}{15\pi^2}\frac{m^{3/2}}{\hbar^3}(g_{\uparrow\uparrow} n_{\uparrow})^{5/2} f\left(\frac{g_{\uparrow\downarrow}^2}{g_{\uparrow\uparrow}g_{\downarrow\downarrow}},\frac{g_{\downarrow\downarrow} n_\downarrow}{g_{\uparrow\uparrow} n_{\uparrow}}\right) 
\label{eq:ExplicitMixtureLHY}
\end{equation}
with $f(x,y)=\sum_{\pm}\left(1+y\pm \sqrt{(1-y)^2+4xy}\right)^{5/2}/(4\sqrt{2})$.

Close to the mean-field collapse condition, the main contribution to the LHY energy comes from the most energetic modes of sound velocity $c_+\propto \sqrt{g}$, which remain hard and give a sizable contribution to the LHY term $\mathcal{E}_{\mathrm{LHY}}\propto c_+^5$. The least energetic modes have a sound velocity $c_-\propto \sqrt{\delta g}$ and become soft when $\delta g\rightarrow 0$. Their contribution to the LHY energy is therefore negligible. Actually, for $\delta g<0$ they even give a small imaginary contribution, which is neglected within this theoretical framework. We will come back to this point in Sec. \ref{sec:ConclusionDroplet}. At the beyond mean-field level, the key difference between the single-component Bose gas and the Bose-Bose mixture therefore comes from the existence of the highest-energy Bogoliubov branch.

The expressions of the mean-field (Eq. \eqref{eq:EMFsoft}) and LHY energy densities (Eq. \eqref{eq:ExplicitMixtureLHY})  explicitly show that in a Bose-Bose mixture the mean-field and beyond-mean-field terms depend on different coupling constants: $\delta g$ for the mean-field term, which vanishes at the mean-field collapse threshold, and $g_{\uparrow\uparrow},\,g_{\downarrow\downarrow}$, and $g_{\uparrow\downarrow}$ for the LHY term, which are comparable and $\sim g$, remain large and give rise to a sizable beyond-mean-field contribution. In other words, in a Bose-Bose mixture the values of the interspin and intraspin interactions can be chosen to cancel the overall mean-field energy, but this does not cancel the quantum fluctuations, making them play a dominant role. This idea was first put forward by D.~S.~Petrov, who realized that quantum fluctuations would stabilize the gas beyond the mean-field collapse threshold, giving rise to a new quantum liquid phase \cite{PetrovPRL2015}. 

\subsection{Stabilization of quantum liquid droplets}\label{subsec:Droplets}

A liquid is commonly understood as a dense phase of matter formed by particles whose positions have short-range correlations but do not display any long range order. A key property of liquids is that they  maintain a fixed volume but lack rigidity to shear stress, and therefore can flow and adopt the shape of their containers. Our understanding of them builds on van der Waals' pioneering work. In 1873, he understood that liquids were formed by discrete particles interacting through an interparticle potential that is repulsive at short distances and attractive at large ones, thus having a minimum at a finite value of the interparticle distance. In classical fluids, the attraction stems from the van der Waals force between induced dipoles, and the repulsion from the Pauli exclusion principle between the electrons of the external shells of the atoms. In this situation, liquids are intrinsically dense, as the equilibration of both forces occurs at distances on the order of the range of the potential.

D.~S.~Petrov proposed in 2015 that a very different type of liquid, ultradilute and independent from the details of the potential, could emerge in an ultracold Bose-Bose mixture due to the competition between the mean-field and beyond-mean-field terms. To understand the origin of this ultradilute liquid origin, let's consider again this system and focus on the situation where its overall mean-field energy is attractive, $\delta g<0$. Close to the collapse condition, and once the ratio of the spin densities has been fixed to the value that minimizes the spin energy of the system  $n_{\uparrow}/n_{\downarrow}=\sqrt{g_{\downarrow\downarrow}/g_{\uparrow\uparrow}}$, the energy density including the beyond mean-field contribution can be written in the following form 
\begin{equation}
\mathcal{E}= -\frac{1}{2}\gamma_{\mathrm{MF}} n^2+\frac{2}{5}\gamma_{\mathrm{LHY}} n^{5/2}.\label{eq:Edroplet}
\end{equation}
Here $n=n_{\uparrow}+n_{\downarrow}$ is the total density and the values of $\gamma_{\mathrm{MF}}$ and $\gamma_{\mathrm{LHY}}$  can be read from Eqs. \eqref{eq:EMFsoft}  and \eqref{eq:ExplicitMixtureLHY}, respectively.

Equation \eqref{eq:Edroplet} explicitly shows that in these conditions the mean-field energy contribution is attractive and scales with $\gamma_{\mathrm{MF}}\propto\left|\delta g\right|$. In contrast, the beyond mean-field LHY contribution is always repulsive and scales with the values of the interspin and intraspin scattering lengths such that $\gamma_{\mathrm{LHY}}\propto g^{5/2}$. Because both terms have opposite signs, they can balance to form a self-bound liquid. Explicitly, this will occur when the system has a pressure $P=0$. Using the thermodynamical relation
$P=n d\mathcal{E}/d n-\mathcal{E}$
we see that the equilibration will take place when the density reaches the value 
\begin{equation}
n_{\mathrm{eq}}=\left(\frac{5\gamma_{\mathrm{MF}}}{6 \gamma_{\mathrm{LHY}}}\right)^2\propto \frac{\delta g^2}{g^5}.   
\end{equation}

The different density scaling of the mean-field and LHY energies is crucial to make the stabilization of a liquid phase robust. Indeed, as soon as $\delta g<0$, the system starts to collapse, since the long-wavelength density modes of wavevector $k\sim\sqrt{m|\delta g| n}$ become imaginary. The corresponding density increase makes the energy of the spin modes grow, and consequently increases the zero-point energy of the system, as $\mathcal{E}_{\mathrm{LHY}}\propto n^{5/2}$. The stronger density-dependence of the repulsive LHY term with respect to the attractive mean-field one ensures that the system can always equilibrate, and no fine-tuning of the interaction strengths is needed. Curiously, the system stabilizes because the large wavelength mean-field instability is blocked by the effect of quantum fluctuations, which take place at short wavelengths \cite{PetrovPRL2015}.

The self-bound state formed by the mixture in the regime of mean-field collapse is known as a quantum liquid droplet. This name, coined by D.~S.~Petrov in his seminal article, reflects the fact that the system in this regime has the key properties of a liquid even though it is not dense. To study such properties, the standard approach is to describe the mixture with an effective low-energy theory: an extended Gross-Pitaevskii equation (eGPE) with an additional repulsive term. It includes the beyond-mean-field effects as an effective potential for the low-energy degrees of freedom of the system, an approach that is valid in the absence of spin excitations. This condition is fulfilled by assuming identical spatial modes for the two components $\Psi_{\uparrow}=\sqrt{n_{\uparrow}}\phi$ and $\Psi_{\downarrow}=\sqrt{n_{\downarrow}}\phi$ with $n_{\uparrow}/n_{\downarrow}=\sqrt{g_{\downarrow\downarrow}/g_{\uparrow\uparrow}}$, which leads to the following eGPE equation
\begin{equation}
i\hbar \partial_t\phi=\left[-\frac{\hbar^2}{2m}\nabla^2+V_{\mathrm{trap}}(r)+\gamma_{\mathrm{MF}} |\phi|^2+\gamma_{\mathrm{LHY}}|\phi|^3\right]\phi.\label{eq:EGPE}
\end{equation}
This equation is analogous to the Gross-Pitaevskii equation describing a single-component BEC Eq. \eqref{eq:GPE}, but with a renormalized scattering length and an additional LHY term. From it, it is straightforward to obtain the main properties of the liquid phase. 

\begin{figure}[b]
\centering
\includegraphics[scale=1]{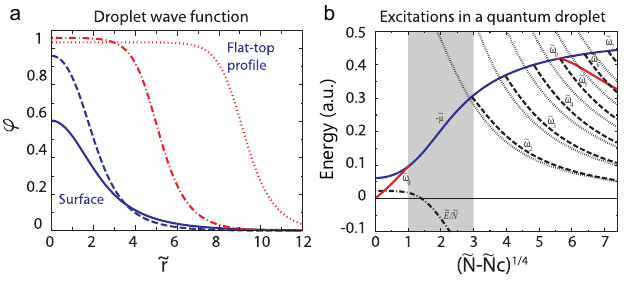}
\caption{\textbf{Properties of the self-bound quantum liquid droplets in a mixture of Bose-Einstein condensates.} (a) Droplet wavefunction $\phi$ as a function of the radial coordinate in renormalized units $\tilde{r}$.  For sufficiently large atom numbers the wavefunction in the center of the droplet reaches a saturation value, leading to a flattop density profile characteristic of a liquid. For small atom numbers, its shape is dominated by surface effects. (b) Excitation spectrum of a droplet as a function of the renormalized atom number $\tilde{N}$ compared to the critical one $\tilde{N_{\mathrm{c}}}$. The blue solid line $-\tilde{\mu}$ corresponds to the particle emission threshold that separates continuum and discrete excitations. At low atom numbers, the droplet has a monopole mode. At high atom numbers, the system exhibits surface ripplon excitations characteristic of a classical liquid. In the intermediate atom number regime, there is a region with no excitations where the system is expected to self-evaporate to zero temperature. Figure extracted from \cite{CabreraPhD2018}, adapted from \cite{PetrovPRL2015}.} 
\label{fig:DropletProperties} 
\end{figure}

First of all, it forms self-bound states, reminiscent of the drops formed by a classical liquid. The density profile of a quantum liquid droplet, depicted in Fig. \ref{fig:DropletProperties}a, is spatially isotropic, reflecting the isotropy of the contact interactions that describe the mixture. For sufficiently large atom numbers, it displays a plateau at its center, with a density corresponding to the equilibrium density $n_{\mathrm{eq}}$, and wings where the density drops to zero over a distance corresponding to the density healing length. The central density is saturated, and increasing the number of atoms in the droplet simply makes its size grow. This behavior is characteristic of the very low compressibility of a liquid phase. 

Moreover, the self-bound droplet is only stable above a critical atom number $N_{\mathrm{c}}$. This is due to the role of kinetic energy in the system, which is a repulsive quantum pressure with a different scaling with atom number $\mathcal{E}_{\mathrm{kin}}\propto N$ compared to the interaction terms $\mathcal{E}_{\mathrm{MF}}\propto N^2$ and $\mathcal{E}_{\mathrm{LHY}}\propto N^{5/2}$. Upon decrease of the atom number, its effect becomes increasingly important and below a threshold $N<N_{\mathrm{c}}$ it is sufficient to dissociate the droplet. Indeed, the shape of small droplets just above the dissociation threshold is mostly given by quantum pressure and surface effects.

Another property which reflects the liquid character of the system is the excitation spectrum of the droplet, depicted in Fig. \ref{fig:DropletProperties}b. At large atom numbers, it displays a succession of surface excitations analogous to the ripplons of liquid helium, which depend on the surface tension of the droplet.  Finally, at lower atom numbers closer to $N_{\mathrm{c}}$, there is a regime where no excitations exist in the system below the particle emission threshold that separates the discrete and continuum parts of the spectrum. In this regime, exciting the droplet should be equivalent to spilling of particles from it. That is, it should self-evaporate to zero temperature. This is an intriguing behavior that would be exciting to explore, but that has until now defied experimental observations.

Although these self-bound droplets share many properties with a conventional liquid, they occur in extremely dilute systems (with densities eight orders of magnitude lower than liquid helium, and still five orders of magnitude lower than an ideal gas). Due to such low density, they have a featureless structure factor and constitute a liquid that goes beyond the standard van der Waals paradigm. Their properties do not depend on the range of the potential but just on the three scattering lengths that describe the system, which makes them universal \cite{PetrovNatPhys2018}. Their existence provides a striking demonstration of the importance of quantum fluctuations, which remain significant even in the weakly interacting regime where the gas parameter satisfies $n a^3\ll 1$. 

\subsection{Competing interactions and quantum liquid droplets in dipolar quantum gases.}
Quantum liquid droplets were predicted in quantum mixtures. However, they were first observed, unexpectedly, in Bose-Einstein condensates of highly magnetic atoms, specifically of the lanthanides dysprosium \cite{KadauNature2016, Ferrier-BarbutPRL2016} and erbium \cite{ChomazPRX2016}. Such systems are described by two interaction terms of very different origins. On the one hand, there is the standard contact interaction of scattering length $a$, which is isotropic and is set to repulsive values in the experiments to ensure the stability of the system. On the other hand, there is the dipole-dipole interaction between the magnetic dipoles associated to the atoms, which is anisotropic and takes the form 
\begin{equation}
V_{\mathrm{dd}}(\mathbf{r}-\mathbf{r'})=\frac{\mu_0\mu_{\mathrm{m}}^2}{4\pi}\frac{1-3\cos^2\theta}{\left|\mathbf{r}-\mathbf{r'}\right|^3},
\end{equation}
where $\mathbf{r}$ and $\mathbf{r'}$ are the position of the dipoles, $\mu_0$ is the magnetic permeability of vacuum, $\mu_m$ is the magnetic dipole of the atoms, and $\theta$ is the angle between the direction of the dipoles and the $z$ axis. This interaction can take repulsive or attractive values depending on whether the dipoles are aligned side to side or head to tail, respectively. Therefore, even if the system has a single component, the competition between both interaction terms can cancel at the mean-field level, making the effect of quantum fluctuations apparent.

The relative strength of the two mean-field interactions is commonly characterized by the parameter $\epsilon_{\mathrm{dd}}=a_{\mathrm{dd}}/a$, where $a_{\mathrm{dd}}$ is the lengthscale of the dipolar interaction $a_{\mathrm{dd}}=m\mu_0\mu_{\mathrm{m}}^2/(12\pi\hbar^2)$. It plays a role analogous to the parameter $\delta a$ in the mixture case. The strength of the LHY term in the dipolar case takes the form  
\begin{equation}
g_{\mathrm{dLHY}}=\frac{32}{3\sqrt{\pi}}g\sqrt{a^3}Q(\epsilon_{\mathrm{dd}})\simeq\frac{32}{3\sqrt{\pi}}g\sqrt{a^3}\left(1+\frac{3}{2}\epsilon_{\mathrm{dd}}^2\right),
\end{equation}
where $Q_5(\alpha)=1/2\int_0^{\pi}\mathrm{d}\theta \sin\theta[1+\epsilon_{\mathrm{dd}}(3\cos^2\theta-1)]^{5/2}$ describes the average angular contribution of the dipolar interactions \cite{LimaPRA2011}. The dipole-dipole interaction is anisotropic. As a result, dipolar quantum droplets are also anisotropic, with an elongated shape along the magnetic field direction that aligns the magnetic dipoles. Analogously to the mixture droplets, they can be described with an extended Gross-Pitaevskii equation which in this case includes a mean-field dipolar interaction term.

Finally, note that quantum gases with dipole-dipole interactions can also be obtained by exploiting Bose-Einstein condensates of polar molecules, where the dipoles are electric instead of magnetic. The dipole-dipole interaction is much stronger in the molecular case \cite{ZhangArXiv2025} and, as a result, it is possible to go beyond the weakly interacting regime. 

\subsection{Experiments with quantum liquid droplets} 
As explained above, the first experimental observation of quantum liquid droplets was done in dipolar Bose-Einstein condensates of the highly magnetic $^{164}$Dy atom \cite{KadauNature2016}. Since then, quantum droplets have been realized in dipolar systems made of $^{164}$Dy \cite{KadauNature2016,Ferrier-BarbutPRL2016, Ferrier-BarbutJPB2016, SchmittNature2016, Ferrier-BarbutPRL2018}, $^{162}$Dy \cite{BoettcherPRR2019}, and $^{166}$Er atoms \cite{ChomazPRX2016}. The original scenario of Bose-Bose mixtures was first investigated by our group at ICFO using spin mixtures of $^{39}$K atoms \cite{CabreraScience2018, CheineyPRL2018}, quickly followed by the LENS group \cite{SemeghiniPRL2018, FerioliPRL2019}. Subsequently, mixture droplets have also been realized in heteronuclear mixtures of $^{41}$K-$^{87}$Rb \cite{DErricoPRR2019,CavicchioliPRL2025} and $^{23}$Na-$^{87}$Rb \cite{GuoPRR2021}. More recently, quantum droplets have been realized in Bose-Einstein condensates of $^{23}$Na-${^{133}}$Cs polar molecules \cite{ZhangArXiv2025}. Below, we review the main experimental observables investigated in the experiments. The focus is mostly on the mixture droplets, but dipolar experiments are summarized as well.

\subsubsection{Self-bound character, critical atom number and liquid-to-gas transition}
\begin{figure}[t]
\centering
\includegraphics[scale=1]{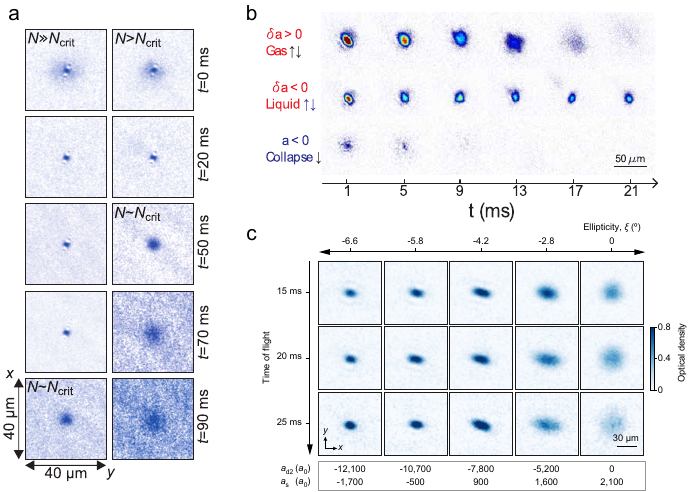}
\caption{\textbf{Self-bound character of quantum liquid droplets in the different systems.} In all cases, the external trapping potential is removed at $t=0$, and the system evolves freely in the absence of confinement afterwards. (a) Dysprosium bosonic atoms with a large magnetic dipole moment. (b) Mixture of potassium Bose-Einstein condensates with competing mean-field interactions of overall strength $\delta a$ (top and middle rows), and single-component BEC of scattering length $a$ (bottom row). In this experiment, a vertical confinement is maintained along the vertical direction to prevent the atoms from falling out of the depth of focus of the imaging system. The self-bound character is therefore only probed along the transverse directions. (c)  Ground state polar NaCs molecules, with large electric dipole characterized by the molecular dipolar length $a_{\mathrm{d}2}$ (see \cite{ZhangArXiv2025}). Figure adapted from \cite{SchmittNature2016,CabreraPhD2018,ZhangArXiv2025}.} 
\label{fig:Self-boundExperiments} 
\end{figure}

The most striking feature of the quantum liquid phase is the fact that it forms self-bound liquid droplets, which do not expand in the absence of external confinement. This behavior has been directly probed in the three systems: atomic mixtures, atomic magnetic dipoles, and molecular electric dipoles. Exemplary images taken in the experiments are displayed in Fig. \ref{fig:Self-boundExperiments}. They show how, upon release of the external confinement, the system does not expand, remaining self-bound. 

\begin{figure}[t]
\centering
\includegraphics[scale=1]{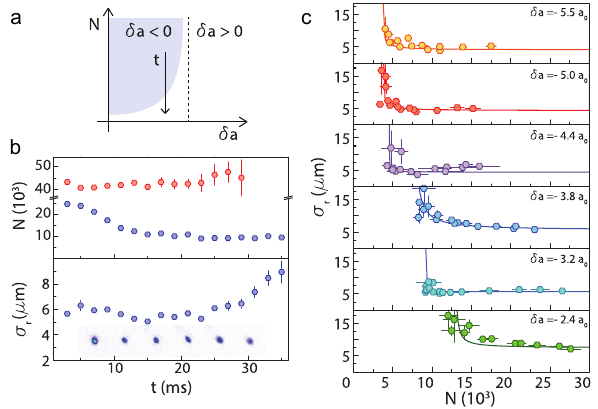}
\caption{\textbf{Determination of the liquid-to-gas phase transition exploiting three-body losses.} (a) Schematic phase diagram of mixture droplets in the $N$ vs. $\delta a$ parameter space. The blue region corresponds to the liquid phase, and the white one to the gas phase. The arrow indicates the evolution of an experimental system originally prepared in the liquid state, but which reduces its atom number over time due to three-body recombination until crossing the phase transition. (b) Evolution of the atom number $N$ (top panel) and transverse size of the system $\sigma_r$ (bottom panel) vs. time $t$. In the liquid phase, $N$ decreases due to three-body losses. It reaches a constant value when the system enters the gas phase, as its size starts to increase and the corresponding density decrease stops the three-body losses. (c) $\sigma_r$ vs. $N$ for various interaction strengths $\delta a$. The sudden increase in system size marks the critical atom number $N_{\mathrm{c}}$ for which the liquid-to-gas phase transition takes place. Figure adapted from \cite{CabreraScience2018}.} 
\label{fig:Liquid-to-gas} 
\end{figure}

As discussed in Sec. \ref{subsec:Droplets}, quantum liquid droplets only remain self-bound down to a critical atom number $N_{\mathrm{c}}$. Below $N_{\mathrm{c}}$, the repulsive effect of quantum pressure is able to dissociate them, leading to an expanding gas. The value of the critical atom number $N_{\mathrm{c}}$ can be measured in experiments using a fairly straightforward method pioneered by the Stuttgart group \cite{SchmittNature2016}, and depicted in Fig. \ref{fig:Liquid-to-gas} for the ICFO mixture droplet experiments. It makes use of the fact that these systems suffer from considerable three-body losses in the gas phase due to their high densities (typical experimental values are on the order of $10^{14}$ atoms/cm$^3$). Therefore, once a self-bound liquid droplet is formed in the absence of confinement, its atom number $N$ quickly decreases over time. When it reaches the critical value $N_{\mathrm{c}}$, the droplet unbinds and the gas starts to expand, quickly decreasing its density and effectively stopping the atomic losses. As a result, the atom number reaches a constant value, corresponding to the critical atom number, see Fig. \ref{fig:Liquid-to-gas}a. Plotting the size of the system as a function of the atom number, as in Fig. \ref{fig:Liquid-to-gas}b, is a particularly clear way of revealing the transition, which appears as an increase in system size at a constant atom number. 

The critical atom number for the phase transition has been measured accurately for both dipolar and mixture droplets. Interestingly, discrepancies with the simple beyond-mean-field picture described in these lecture notes are present in several of these experiments \cite{SchmittNature2016, CabreraScience2018, BoettcherPRR2019}. Since mean-field interactions have been practically canceled out, only the small effect of quantum fluctuations remains, which makes the system sensitive to other small energy corrections.
Examples are the exact form of the dipole-dipole interaction, the effective range of the contact interaction potential, or the exact form of the beyond-mean-field corrections in the presence of confining potentials, which deserve further theoretical investigation as discussed in Sec. \ref{sec:ConclusionDroplet}. A precise theoretical description of molecular quantum droplet experiments beyond the weakly interacting regime will be even more challenging. 

\subsubsection{Droplet collisions}
\begin{figure}[t]
\centering
\includegraphics[scale=1]{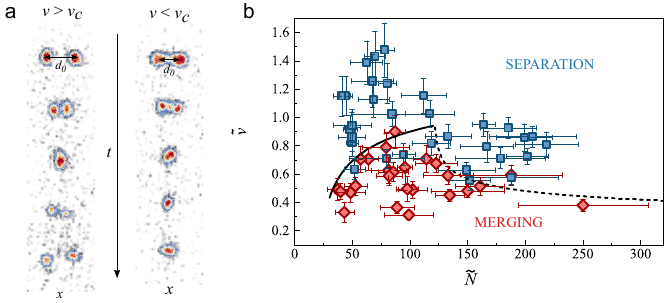}
\caption{\textbf{Collision between mixture droplets.}  (a) Exemplary images of binary droplet collisions. For low relative velocities the two droplets merge into a single one after the collision. Above a critical velocity $v_c$, they bounce against each other, separating after the collision. (b) Observed outcome of the collision (separation or merging) as a function of the relative velocity and the atom number. Both quantities are renormalized to dimensionless units,  $\tilde{v}$ and $\tilde{N}$. The lines indicate the expected critical velocity obtained from the liquid drop model discussed in the text. They allow distinguishing the regimes where the droplets are compressible (solid line, at low atom numbers) from the one where they become closer to incompressible and surface tension effects play an important role on the collision dynamics (dashed line, at higher atom numbers). Figure adapted from \cite{FerioliPRL2019}. } 
\label{fig:DropletCollisions} 
\end{figure}

A powerful method to gain more information on the properties of quantum liquid droplets is to make them collide. In the case of mixture droplets, droplet collisions were studied by the LENS group \cite{FerioliPRL2019}. In the experiments, they created two droplets of well controlled relative velocity $v$ and investigated the outcome of the collision as a function of $v$ and of the droplet atom number. Analogously to what occurs in classical liquids, the collision of two droplets can have two different outcomes. For relative velocities below a critical value $v_c$, the two droplets merge into a single larger one. In contrast, for $v>v_c$, the droplets bounce against each other and re-emerge after the collision as two distinct objects. Figure \ref{fig:DropletCollisions}a shows exemplary experimental images of the two scenarios. While the value of $v_c$ that separates the merging and separation scenarios depends on the mean atom number of the droplets at the time of the collision, the functional form is very different in the regime of small and large droplets, see Fig. \ref{fig:DropletCollisions}b. Within a liquid drop model, this can be understood with a simple scaling argument. The possibility of forming a single droplet during the collision is related to the capability of the resulting merged droplet to absorb the excess kinetic energy. For large droplets, with distinct bulk and surface, the energy can be absorbed by the excitation modes of the surface discussed in Sec. \ref{subsec:Droplets} and the scaling is $v_c\sim N^{-1/6}$. For small droplets, whose shape is dominated by surface effects, the relevant energy scale is instead the overall binding energy of the droplet and $v_c$  increases with atom number. The change of scaling of the critical velocity separating merging and separation of droplets is therefore a sensitive probe of the crossover from compressible to nearly incompressible droplets. Collisions between dipolar droplets were investigated by the Stuttgart group as well \cite{Ferrier-BarbutJPB2016}. However, due to the repulsive interactions existing between the droplets in this case, only bouncing behavior is observed.

\subsubsection{Interplay with a harmonic trap and soliton-to-droplet transition}
\begin{figure}[t]
\centering
\includegraphics[scale=1]{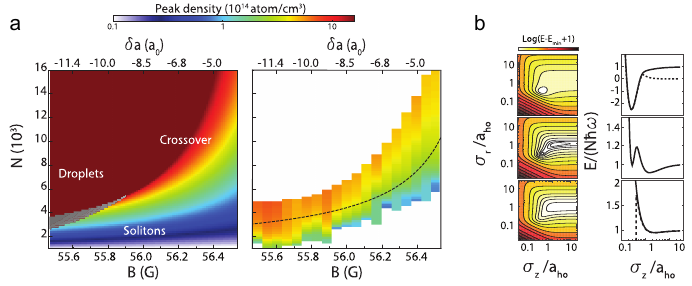}
\caption{\textbf{ Soliton-to-droplet transition of a Bose-Bose mixture in an optical waveguide.}  (a) Soliton-to-droplet phase diagram depicting the peak density of the self-bound solution as a function of atom number $N$ and effective scattering length $\delta a$ (controlled through the value of the external magnetic field $B$). For large attraction, droplets and solitons are distinct solutions that coexist in a bistable region (grey area). For weak attraction, there is only one self-bound state that evolves smoothly from soliton-like at low atom number to droplet-like at high atom number. The theoretical predictions (left panel) are in good agreement with the experimental observations (right panel). The dashed line represents the mean-field collapse threshold. (b) Predicted self-bound states in the strongly attractive region, where droplets (top row) and solitons (bottom row) coexist in a bistable region (central row). Figure adapted from \cite{CheineyPRL2018}.} 
\label{fig:Soliton-to-droplet} 
\end{figure}

The experiments described above were performed in the absence of trapping potentials, as they aimed at probing the self-bound character of the quantum droplets. The addition of a harmonic trap brings interesting new aspects, in particular in low-dimensional systems. At ICFO we investigated this topic by trapping a Bose-Bose mixture with competing interactions in a one-dimensional waveguide \cite{CheineyPRL2018}. In this geometry, a conventional Bose gas with attractive interactions forms bright solitons: self-bound states that are stabilized by the competition of the attractive mean-field interaction term and the repulsive effect of quantum pressure. They are however only stable up to a critical atom number, above which the system becomes effectively three-dimensional and mean-field collapse takes place. In an attractive Bose-Bose mixture, both quantum droplets and bright solitons are stable. Therefore, as soon as the overall mean-field energy is attractive, a self-bound state always exists. Its nature depends on the interaction strength $\delta a$ and the atom number $N$. For weak attraction, the soliton and droplet regimes are smoothly connected: the system supports a single self-bound state whose nature evolves from droplet-like at large $N$ to soliton-like at small $N$. In contrast, for strong attraction we find two distinct behaviors: a high-density solution (of peak density as large as $n\sim10^{16}$ atoms/cm$^3$) for large $N$, and a low-density one ($n\sim10^{13}$ atoms/cm$^3$) for small $N$. In between, there is a bistable regime where both solutions are possible. As depicted in Fig. \ref{fig:Soliton-to-droplet}, this situation was not only investigated theoretically, but also experimentally. Subsequently, a similar scenario has been predicted for dipolar gases, but still needs to be realized experimentally. Interestingly, since dipolar BECs can host solitons also in two dimensions, the soliton-to-droplet transition in this case is expected to occur also in 2D \cite{Ferrier-BarbutPRA2018, NataleCommunPhys2022, SchubertArXiv2026}.

\subsection{Experiments in the gas phase} 

\begin{figure}[t]
\centering
\includegraphics[scale=1]{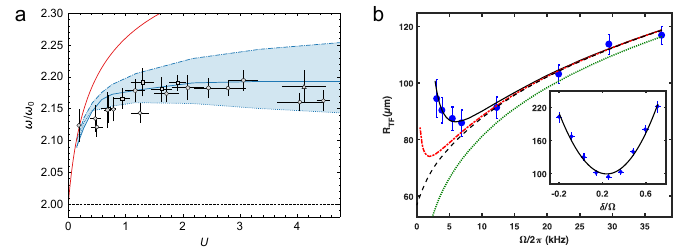}
\caption{\textbf{Quantum fluctuation effects in the gas phase.} (a) The equation of state of a bosonic mixture when approaching the liquid phase is modified by quantum fluctuations. Its effect can be probed by measuring the frequency of the breathing mode as a function of the interaction parameter  , where $a_{\mathrm{ho}}=\sqrt{\hbar/m\omega_0}$ is the harmonic oscillator length and $\omega_0$ the trap frequency. The stiffening of the frequency above the standard $2\omega_0$ result is a directly probe of the effect of quantum fluctuations. (b) Effect of the Lee-Huang-Yang energy on the expansion of the lowest-energy dressed state of a coherently-coupled gas where the effective mean-field term cancels. The expansion is driven by the LHY term, which can be controlled through the Rabi frequency of the coupling field. At low Rabi frequencies, it has been predicted to yield effective three-body interactions. However, this regime is currently complex to reach experimentally due to magnetic field fluctuations, which cause an occupation of the highest-dressed state and are the reason for the increase in size observed experimentally for the smallest values of $\Omega$. Figure adapted from \cite{SkovPRL2021, LavoinePRL2021}.} 
\label{fig:LHYgas} 
\end{figure}

The existence of the self-bound liquid phase is the most spectacular demonstration of the relevance of quantum fluctuations in systems with competing interactions. However, in the vicinity of the liquid-to-gas transition, quantum fluctuations play an important role also in the gas phase. 
In mixture systems, the Aarhus group investigated their effect on the equation of state of the system by measuring the frequency of its breathing mode for both $\delta a<0$ and $\delta a >0$ \cite{SkovPRL2021}. As depicted in Fig. \ref{fig:LHYgas}a, the experiment observes an increase of the breathing mode frequency above the mean-field prediction $\omega/\omega_0=2$, providing clear evidence for the Lee-Huang-Yang energy term \cite{JorgensenPRL2018}. A precise interpretation of the experiments requires however the incorporation of three-body losses. The excitation modes of a dipolar system in the crossover between a Bose-Einstein condensate and a quantum droplet have also been investigated by the Innsbruck group. Already in their very first droplet experiments they observed a clear increase of frequency of the lowest-lying quadrupole mode when entering the droplet regime, as expected theoretically \cite{ChomazPRX2016}. In recent experiments, they have exploited instead Bragg spectroscopy to confirm the generalized stiffening of excitations when approaching the liquid phase \cite{HouwmanPRL2024}. 

A different class of experiments on the gas phase was performed by our group and the Palaiseau group. It consists on subjecting a $^{39}$K gas with competing interactions to a coherent coupling between the two components using a radio-frequency (rf) field of Rabi frequency $\Omega$ and detuning $\delta$. The resulting system can be understood as a mixture of dressed states separated by an energy gap $\hbar\Omega$ and with renormalized 
scattering properties. These effective interactions 
% interaction properties, which
depend on the scattering lengths of the two coupled spin components and can be flexibly controlled through the parameters of the rf field \cite{NicklasPRL2011, ShibataPRA2019, SanzPRL2022}. While the highest-energy dressed state suffers from inelastic dressed-state changing collisions, the lowest-energy one has long lifetimes \cite{SanzPRL2022}. When only 
the lowest 
% that 
dressed state is occupied, an effective single-component description of the system is appropriate, and due to the competing interactions of the original system, its effective scattering length can be set to zero. However, unlike in a true single-component gas, the role of quantum fluctuations remains sizeable in this situation. This is due to the fact that the highest-energy dressed state still contributes to the Lee-Huang-Yang term. Interestingly, the density scaling of the resulting Lee-Huang-Yang term changes with $\Omega$. At large values, the main effect is to renormalize the value of the two-body mean-field term, as the Palaiseau group observed experimentally \cite{LavoinePRL2021}, see Fig. \ref{fig:LHYgas}(b). At low values, it takes the form of an effective three-body term, a regime that is very challenging to reach experimentally due to magnetic field fluctuations, but which should become accessible in future experiments.  

\subsection{Conclusion and outlook}\label{sec:ConclusionDroplet}

Ten years have passed after the theoretical prediction of quantum liquid droplets stabilized by quantum fluctuations in Bose-Bose mixtures \cite{PetrovPRL2015}, and almost as long since their first experimental observation in dipolar quantum gases \cite{KadauNature2016, Ferrier-BarbutPRL2016, SchmittNature2016, ChomazPRX2016} and mixtures of BECs \cite{CabreraScience2018, CheineyPRL2018, SemeghiniPRL2018}. The research area of ultradilute quantum liquids in Bose-Einstein condensates with competing interactions is now firmly established. The main properties of such universal liquids, in particular their self-bound character and the liquid-to-gas transition, have been investigated in a number of experiments \cite{SchmittNature2016, CabreraScience2018, SemeghiniPRL2018, GuoPRR2021}. The difference with other self-bound states, such as bright solitons, has been clarified \cite{CheineyPRL2018}, and experiments studying droplet collisions and the formation dynamics of multiple droplets have provided indirect evidence of their surface tension \cite{Ferrier-BarbutJPB2016, FerioliPRL2019, CavicchioliPRL2025}. Moreover, the equation of state of the gas phase in the regime of mean-field collapse has also been explored through measurements of its collective excitations \cite{ChomazPRX2016, SkovPRL2021}, expansion dynamics \cite{LavoinePRL2021}, and Bloch oscillations \cite{NataleCommunPhys2022}, providing clear evidence of the Lee-Huang-Yang energy term.

However, many properties of the system remain open. First of all, while the beyond-mean-field theory proposed by D.~S.~Petrov qualitatively predicts the observed quantum droplet properties, precise measurements of the liquid-to-gas phase transition have revealed quantitative differences with theory whose origin still needs to be clarified \cite{SchmittNature2016, CabreraScience2018, BoettcherPRR2019}. A limitation of Petrov’s theory is that it ignores the complex part of the Bogoliubov spectrum when computing the beyond mean-field correction.   Another important point is that the Lee-Huang-Yang term is typically included in the description through a local density approximation, which may not be that accurate in long-range interacting systems or when the system is dense \cite{BoettcherPRR2019}. Finally, the cancellation of mean-field energy terms in systems with competing interactions certainly makes quantum fluctuation effects apparent, but they remain relatively small. As a result, other small corrections to the energy of the system, such as the exact form of the dipole-dipole interactions \cite{OldziejewskiPRA2016} or finite range corrections to the contact interaction term \cite{CikojevicNJP2020}, might become relevant as well. Their inclusion may improve the comparison with the experiments, but this would imply that the liquid phase is not completely universal after all. 

A more fundamental limitation of current quantum droplet theories is that they are intrinsically limited to weak interactions and small gas parameters. While this was an excellent approximation in the first experiments, the recent polar molecule experiments can definitely go beyond this paradigm \cite{ZhangArXiv2025}. There have therefore been significant efforts to overcome these limitations and improve the theoretical description of quantum droplets. A broad range of analytical approaches have been discussed in literature, including the investigation of quantum fluctuations at the dimensional crossover \cite{EdlerPRL2017,IlgPRA2018, ZinPRA2018}, the use of diagrammatic Beliaev techniques \cite{GuPRB2020}, of the hypernetted-chain Euler-Lagrange method \cite{HebenstreitPRA2016, StaudingerPRA2018}, of Gaussian-state theory including squeezing effects \cite{ShiArXiv2019, WangPRR2020}, the inclusion of bosonic pairing \cite{HuPRL2020, HuPRA2020} or of higher order corrections to the Bogoliubov speed of sound \cite{OtaSciPost2020}. Moreover, from the numerical point of view, quantum Monte-Carlo simulations have been used to investigate quantum droplets and their properties \cite{SaitoJPSJ2016, MaciaPRL2016, CintiPRL2017, CikojevicPRB2018, CikojevicPRA2019, BoettcherPRR2019}, showing indeed small deviations with respect to the standard droplet theory \cite{BoettcherPRR2019}. Droplets have also been investigated numerically using DMRG simulations, which allows going beyond the weakly interacting regime, but includes new effects linked to the lattice potential \cite{MoreraPRR2020, MoreraPRL2021}. 

Besides these theoretical aspects, many properties of quantum droplets remain to be investigated experimentally. One of them is their superfluidity, which has not been proved so far. Dipolar experiments investigated the scissors mode \cite{Ferrier-BarbutPRL2018}, which probes the moment of inertia of the system. In standard conditions, the irrotational character of the superfluid yields a moment of inertia that is very different from the rigid-body value and, as a result, the scissors mode frequency provides a clear indication of superfluidity. In dipolar quantum droplets, however, the very anisotropic density distribution results in a moment of inertia that is already close to the rigid-body value, and this mode cannot be used to probe superfluidity. A clear smoking gun for superfluidity would be the nucleation of vortices inside the droplets \cite{KartashovPRA2018}. This requires however droplets with a large bulk region, significantly larger than the vortex size, which is given by the density healing length. Unfortunately, the corresponding equilibrium densities are large, leading to short lifetimes that would make vortex nucleation extremely hard. Proving the superfluidity of quantum droplets remains therefore an open question. 

Another fundamental question is the role of finite temperature on the droplet properties. In the mixture case, the self-evaporation mechanism discussed in Sec. \ref{subsec:Droplets} should be able to automatically create zero-temperature droplets. However, three-body losses have again prevented its observation. In the case of $^{39}$K spin mixtures, their effect is aggravated by the fact that they primarily affect one of the spin components, yielding to the expulsion of particles of the other spin  component in order to maintain the correct ratio of spin densities and greatly complicating the interpretation of potential experiments \cite{FerioliPRR2020}. In dipolar droplets, the effect of a finite temperature might be unavoidable. It has been theoretically proposed that finite temperature should increase the critical atom number of the liquid-to-gas phase transition \cite{BoudjemaaAP2017, AybarPRA2019}, but this effect has not yet been investigated experimentally, and performing droplet thermometry is in itself a formidable challenge. Finally, the investigation of droplets in low-dimensional systems, which has only been investigated theoretically so far \cite{PetrovPRL2016, ParisiPRL2019}, remains a completely open field, and might allow to circumvent several of the challenges raised above.

The field of quantum droplets has evolved however in a different direction. The realization of arrays of dipolar droplets, which form regular structures due to the repulsive nature of droplet-droplet interactions, raised the question of their potential supersolidity. Indeed, these systems break translational symmetry. At the same time, each droplet is very likely superfluid. Therefore, already in the very first experiments it was understood that, if a regime where particle exchange between the droplets could be established, the system would simultaneously display a crystalline structure and phase coherence, and would constitute an excellent candidate for supersolidity \cite{KadauNature2016}. Subsequent experiments searched for it \cite{WenzelPRA2017} and finally manage to show that dipolar droplet arrays could indeed enter a supersolid regime, opening again a new research field \cite{TanziPRL2019, BöttcherPRX2019, ChomazPRX2019}. Bosonic mixtures with tunable interactions can also become supersolid when subjected to spin-orbit coupling \cite{LiNature2017} and share many features with dipolar systems, although in this case the link with quantum droplets is not required. The investigation of supersolidity became therefore the natural evolution of the field of quantum droplets, and is the topic of the next section.

\section{Supersolidity in spin-orbit-coupled Bose-Bose mixtures with repulsive interactions}\label{sec:Supersolid}

Supersolidity is a highly counterintuitive phase of matter that combines the crystalline structure of a solid with the frictionless flow of a superfluid. While the idea of a supersolid dates from the '60s and stems from the field of solid helium, it could only be realized with Bose-Einstein condensates, first of ultracold atoms \cite{RecatiNRP2023} and more recently of exciton-polaritons \cite{TrypogeorgosNature2025, MuszyskiarXiv2024}. In this section, we review the concept of supersolidity and discuss the existing quantum gas platforms that are simultaneously phase coherent and spontaneously break translational symmetry. Among them, dipolar quantum gases have emerged as the reference supersolids in quantum gases. We give a brief review of the key experiments that have been carried out in those systems and their connection with the quantum liquid droplet physics of Sec. \ref{sec:Droplets}. We then discuss supersolidity in Bose-Einstein condensates subjected to spin-orbit coupling. We theoretically show that these systems can be understood as a mixture of Bose-Einstein condensates, which is precisely the topic of these lecture notes, and present the key experiments demonstrating the supersolid nature of the stripe phase of spin-orbit-coupled BECs. We conclude the section by highlighting some exciting research directions on the investigation of spin-orbit-coupled supersolids.

\subsection{Supersolidity in quantum gases}\label{sec:SupersolidIntro}
A supersolid is a phase that spontaneously breaks two continuous symmetries: translational symmetry, which gives it a modulated density profile as in a solid, and U(1) gauge symmetry, which gives it global phase coherence, as in a superfluid. Originally, this situation was speculated to appear in solid helium. There, either the helium atoms themselves (due to their small mass and large zero-point motion) \cite{LeggettPRL1970} or vacancies of the crystal \cite{AndreevJETP1969, ChesterPRA1970}, would delocalize sufficiently to enable superfluid-like flow through the crystal lattice. However, and despite intense experimental efforts, no unambiguous experimental observation of supersolid behavior in $^4$He has been reported so far \cite{BalibarNature2010}. 

\begin{figure}
\centering
\includegraphics[scale=1]{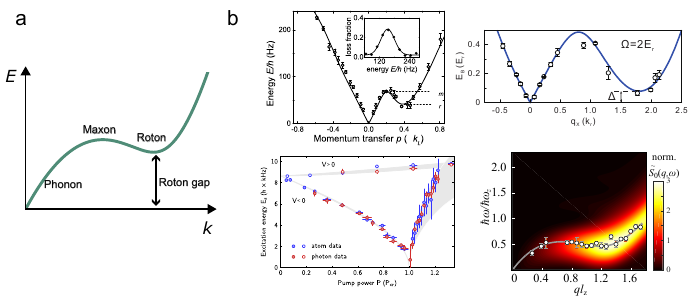}
\caption{\textbf{Roton excitation spectra in different platforms.} (a) Exemplary spectrum with a roton minimum. (b) Experimentally measured roton excitation spectrum in a shaken optical lattice (upper left), in an optical cavity (lower left), in a spin-orbit-coupled BEC (upper right) and in a dipolar BEC (lower right). All four systems show a roton mininum. Figure adapted from \cite{HaPRL2015,MottlScience2012, JiPRL2015,PetterPRL2019}. }
\label{fig:roton} 
\end{figure}

The investigation of supersolidity has therefore moved to other systems, specifically quantum gases. There, supersolidity emerges from the opposite limit, that of a superfluid system which crystallizes spontaneously. Its constituents tend to spontaneously clump at a given lengthscale, giving rise to a modulation of the density of period $d$. On the approach to this transition, the dispersion relation of the atoms displays a minimum at non-zero momentum $k\sim1/d$, see Fig. \ref{fig:roton}a. Its energy reaches zero when entering the supersolid phase. The situation is analogous to that of liquid helium, and therefore the minimum receives the name of roton. Thanks to their large degree of control, Bose-Einstein condensates enable the engineering of such rotonized excitation spectra in a number of platforms: shaken optical lattices \cite{HaPRL2015}, optical cavities \cite{MottlScience2012}, spin-orbit-coupled BECs \cite{KhamehchiPRA2014, JiPRL2015}, and dipolar systems \cite{PetterPRL2019}, as depicted in Fig. \ref{fig:roton}b. Remarkably, the parameter regime where the energy of the roton reaches zero and the superfluid spontaneously crystallizes can be reached in the last three platforms. Bose-Einstein condensates that spontaneously break translational symmetry, showing a modulated density profile in the absence of an externally-imposed periodic potential while maintaining phase coherence have been realized in optical cavities \cite{LeonardNature2017, LéonardS2017, SchusterPRL2020}, in the presence of spin-orbit coupling \cite{LiNature2017, PutraPRL2020}, and in dipolar gases \cite{TanziPRL2019, BöttcherPRX2019, ChomazPRX2019}. Moreover, the coexistence of superfluidity and a crystalline density pattern has also been observed in driven Bose-Einstein condensates where the scattering length is modulated periodically in time \cite{LiebsterPRX2025}.

Are all those systems supersolid? While they have many points in common, the platforms listed above also display significant differences. The stationary patterns of driven quantum gases constitute dynamic far-from-equilibrium states. In contrast, cavity, spin-orbit-coupled and dipolar BECs focus on equilibrium (or near-from-equilibrium) physics. The different systems also differ by some of their excitation modes. Due to the two spontaneously broken continuous symmetries, which are associated to two different order parameters, a supersolid displays two gapless Goldstone modes and a plethora of superfluid and crystal excitations. Remarkably, such excitations can be probed in the different platforms \cite{LéonardS2017, TanziN2019, GuoN2019, NatalePRL2019, TanziScience2021, NorciaPRL2022, BiagioniNature2024, CasottiN2024, LiebsterNatPhys2025} and exploited to characterize the realized states. Among them, some authors put the compressibility of the supersolids’ emergent crystal structure as an additional requirement for full-fledged supersolidity, given that solids naturally host crystal phonons. This property is lacking in BECs with effective long-range interactions induced by single-mode cavities, but can be recovered if multimode cavities are used instead \cite{GuoNature2021}. It has also been observed in dipolar quantum gases \cite{TanziN2019, NatalePRL2019, NorciaPRL2022}, driven quantum gases \cite{LiebsterNatPhys2025}, and spin-orbit-coupled BECs \cite{ChisholmScience2026}. Quantum gases therefore constitute an ideal platform to study supersolid-like states of matter in many different forms. In the following, we focus on supersolidity in dipolar quantum gases and spin-orbit-coupled BECs, which display strong connections with the quantum droplets and quantum mixture physics discussed in Sec. \ref{sec:Droplets}.

\subsection{Dipolar supersolids}\label{sec:dipolarsupersolids}

Although supersolidity was observed earlier in spin-orbit-coupled condensates \cite{LiNature2017}, the platform that has arguably delivered the most extensive investigation of supersolidity is dipolar quantum gases. As there are already several excellent review articles on such systems \cite{BöttcherRPP2021, chomazRoPP2022,  RecatiNRP2023}, we give here only a summary of some marking results. Our goal is to make the connection with droplet physics and provide the necessary context for the investigation of supersolidity in mixtures of spin-orbit-coupled BECs.

\subsubsection{Phase diagram}
\begin{figure}[ht]
\centering
\includegraphics[scale=1]{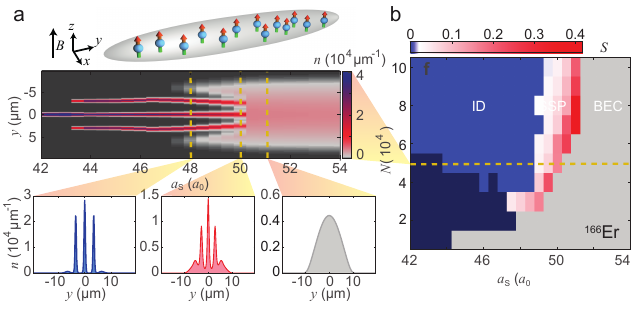}
\caption{\textbf{Phase diagram of a dipolar BEC.} (a) Illustration of the different phases of a dipolar system in a cigar-shaped trap, as depicted on top. The plot shows how the system evolves for different scattering length $a_\mathrm{S}$. In the isolated droplet regime the density drops to zero in between the droplets, whereas in the supersolid regime the density is finite over the whole system but displays a periodic density modulation. In the BEC regime the density modulation disappears. (b) Phase diagram depending on scattering length and particle number. The supersolid phase fills a small region in the phase diagram in between the isolated droplet and BEC phases. The colorbar designates the overlap between neighboring droplets. Figure extracted from \cite{ChomazPRX2019}.}
\label{fig:dipolar-phase-diagram} 
\end{figure}

In the regime of competing dipolar and contact interactions, and in suitable geometries such as elongated optical traps, dipolar droplets can form droplet arrays. These arrays break translational symmetry in their arrangement. However, these systems are a priori not phase coherent and indeed the lack of coherence between independent droplets has been shown experimentally \cite{WenzelPRA2017}. By fine-tuning the system parameters, however, tunneling between the droplets can be restored and phase coherence between the droplets establishes. Figure \ref{fig:dipolar-phase-diagram} illustrates the theoretical phase diagram of a dipolar $^{166}$Er system depending on atom number and contact interactions for a fixed cigar-shaped trap geometry.  Upon increasing the contact interactions, the system evolves from independent droplets into a supersolid regime, which has non-zero density in between the fringes and allows for particle flow between the density peaks.  If contact interactions are increased further, the supersolid fringes dissolve into an unmodulated BEC. This supersolid phase is thus sandwiched in a small region of the phase diagram,  in between the independent droplet regime and the BEC phase.  

\subsubsection{Probing broken translational symmetry and phase coherence}
\begin{figure}[ht]
\centering
\includegraphics[scale=1]{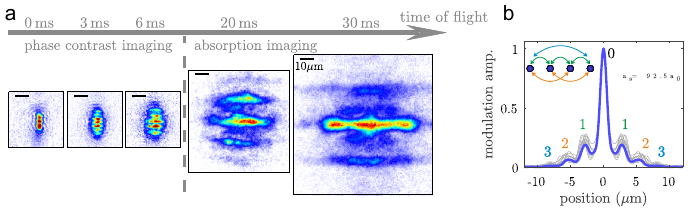}
\caption{\textbf{A dipolar supersolid.} (a) Images of a dipolar supersolid after variable expansion times. For short times the system shows a density modulation. It is imaged using phase-contrast imaging. For longer time-of-flight times the images show the interference pattern of the expanding atoms. They are imaged using absorption imaging. (b) Coherence of the system. The plot shows the Fourier transform of the interference pattern. Grey lines are single images, while the blue line is the averaged signal. The peaks correspond to the coherence between nearest (1), next-nearest (2) and next-next-nearest (3) neighboring fringes. The defined side-peaks with little shot-to-shot variance demonstrate the phase coherence of the system. Figure adapted from \cite{BöttcherPRX2019}.} 
\label{fig:dipolar-supersolid} 
\end{figure}

In 2019, two years after observing the first supersolid in a spin-orbit-coupled BEC \cite{LiNature2017}, a set of experiments in Pisa \cite{TanziPRL2019}, quickly followed by the Innsbruck and Stuttgart groups \cite{ChomazPRX2019,BöttcherPRX2019}, demonstrated supersolidity in dipolar BECs. Figure \ref{fig:dipolar-supersolid} shows images of a $^{162}$Dy dipolar supersolid. The in situ density distribution shows a modulated density profile with three prominent fringes. The complementary time-of-flight data displays the interference pattern of the expanding atoms. The interference peaks demonstrate the phase coherence between neighboring and next-nearest-neighbor fringes. 

\subsubsection{Collective excitations} \label{sec:dipolar_excitations}
With their experimental access to both the phase coherence and the density modulation, dipolar supersolids are in an excellent position to probe supersolid dynamics. As explained in Sec. \ref{sec:SupersolidIntro}, the two broken symmetries, translational symmetry and U(1) gauge symmetry, lead to a rich excitation spectrum including crystal and superfluid excitations. 

\begin{figure}
\centering
\includegraphics[scale=1]{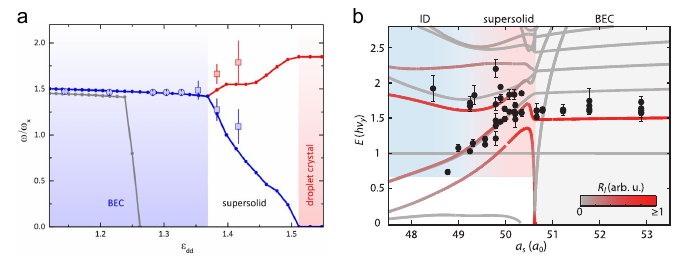}
\caption{\textbf{Excitation frequencies in dipolar gases in the BEC to supersolid to isolated droplets phase diagram.} (a) Response frequencies of a dipolar dysprosium system. Solid lines show numerical simulations of the real-time dynamics using the dipolar eGPE. The lower mode disappears at the transition to the isolated droplet phase (b) Response frequencies of a dipolar erbium system extracted \emph{via} principal component analysis (PCA). Different PCA components correspond to different excited modes and are found at different frequencies.  The solid lines show theory predictions from Bogoliubov-de Gennes theory and the colorbar refers to the strength with which the excitation scheme couples to the mode. Figure adapted from \cite{TanziN2019,NatalePRL2019}.} 
\label{fig:dipolar-excitation-spectrum} 
\end{figure}

First observations on the dynamics in dipolar supersolids showed indeed the appearance of two excitation branches. Figure \ref{fig:dipolar-excitation-spectrum} shows two experiments probing the compressional modes of $^{162}$Dy \cite{TanziN2019} and $^{166}$Er \cite{NatalePRL2019}. 
In the BEC regime, both experiments excite the axial breathing mode, using a sudden change of the scattering length in the dysprosium experiment and a trap excitation scheme in the erbium one. Upon entering the supersolid phase, they observed that the mode splits and several excitations appear with different frequencies. The main excitations correspond to modes with a superfluid and a crystal compression character. The crystal compression mode has also directly been linked to the crystal being able to support phonons \cite{HertkornPRX2021}, similar to a real solid. The appearance of these excitation branches owing to the double symmetry breaking showcases the peculiar nature of supersolid dynamics. 

\begin{figure}[b]
\centering
\includegraphics[scale=1]{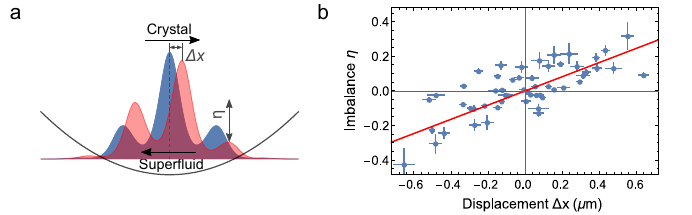}
\caption{\textbf{Low-energy Goldstone mode in a dipolar supersolid.} (a) Illustration of the low-energy Goldstone mode in a harmonic trapping potential. For a three-fringe supersolid, the ground state features a central fringe at the trap center. A phase shift of the crystal is compensated by a superfluid counterflow that keeps the center of mass centered in the trap and leads to an imbalance of the side fringes. (b) Statistical analysis of the fringe displacement $\Delta x$ versus imbalance $\eta$. The data shows a linear dependence (red line) between $\Delta x$ and $\eta$, as expected when the crystal Goldstone mode is excited. Figure adapted from \cite{GuoN2019}.} 
\label{fig:goldstone-dipolar} 
\end{figure}

Another study on dipolar dynamics looked into the crystal Goldstone mode, which is the lowest energy mode of the crystal phonon branch and corresponds to a sliding of the supersolid fringes. In an infinite system this mode is gapless, reflecting the degeneracy with respect to global phase shifts of the density modulation. In a harmonically trapped system, however, the degeneracy is lifted. Due to the envelope of the modulated density, different spatial phases of the crystal correspond to different energies. The energy is minimized for a symmetric pattern with its center of mass in the center of the trap. This means that the Goldstone mode acquires a small but finite energy gap. Furthermore, in a finite system, shifting the crystal pattern leads to a translation of the center of mass.
To keep it in the center of the trap, a particle flow compensates for the phase shift. This redistribution of particles between the fringes leads to an asymmetric pattern. The process is illustrated in Figure \ref{fig:goldstone-dipolar}a. The Stuttgart group used this connection between phase shift and asymmetry of the fringe pattern to study the Goldstone mode \cite{GuoN2019}. A statistical analysis of many repetitions of the experiment, shown in Fig. \ref{fig:goldstone-dipolar}b, revealed a linear dependence between displacement of the central fringe and imbalance of the side fringes. They concluded that the low-energy Goldstone mode is indeed excited in their system and that a superfluid flow exists between the fringes. 

\subsubsection{Supersolidity in two-dimensional systems and vortices}
\begin{figure}[h]
\centering
\includegraphics[scale=1]{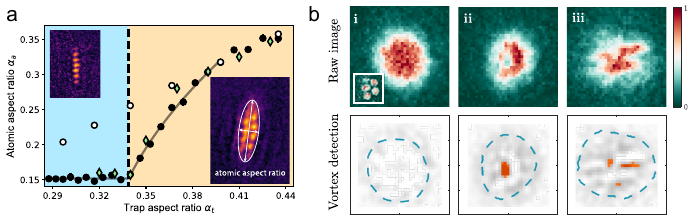}
\caption{\textbf{Dipolar supersolids in two dimensions and vortices.} (a) Phase transition from a one-dimensional to a two-dimensional supersolid, which occurs at a critical trap aspect ratio. Green diamonds indicate dipolar eGPE predictions, white markers show the aspect ratio of an unmodulated BEC. The insets are example images of a one and a two-dimensional supersolid. (b) Detection of vortices in a dipolar supersolid. Panels (i)-(iii) show exemplary images of vortices for increasing rotation strength. For detection the system was quickly ramped from the supersolid to the BEC regime after the rotation. The inset in (i) shows the initial state, a two-dimensional dipolar supersolid. The detected vortices are shown in orange in the bottom row. Figure adapted from \cite{NorciaN2021,CasottiN2024}.} 
\label{fig:2d_dipolar_supersolid} 
\end{figure}

As these experiments demonstrate, dipolar supersolids host an intriguing excitation spectrum including a crystal phonon excitation branch similar to real solids. However, while real solids usually consist of a three-dimensional crystal, supersolids in ultracold quantum gases do not. In the experiments with dipolar atoms revisited so far, the cigar-shaped trap defines an axis along which the translational symmetry is spontaneously broken. By changing the trap geometry, dipolar supersolids have been achieved that break the translational symmetry  in two dimensions \cite{NorciaN2021}. The Innsbruck group observed a phase transition from a density modulation along one direction to a two-dimensional pattern upon changing the aspect ratio of the trap, see Fig. \ref{fig:2d_dipolar_supersolid}a. The transition point depends on the trap geometry as well as the atom number in the system. Two-dimensional dipolar supersolids thus require an increased atom number as compared to the one-dimensional case. 

In such a two-dimensional supersolid system the same group also achieved the observation of quantized vortices \cite{CasottiN2024}. This quantized response to a rotation of the system is considered a smoking gun for superfluidity. Its observation complements previous measurements of the phase coherence \emph{via} interference.  In a supersolid, it is challenging to detect vortices through their density depletion, because they are located within the low density region in between the fringes. To probe their presence, the experiment ramped the scattering length after the rotation, thus passing over the phase transition into the BEC regime. While the topologically protected vortices survive, the density modulation melts, which increases the contrast of the vortices. Figure \ref{fig:2d_dipolar_supersolid}b shows exemplary images containing zero, one and three vortices, and providing an unambiguous proof of the superfluidity of the system.

In conclusion, experiments on dipolar quantum gases have firmly established the broken translational invariance and phase coherence of the system, and have probed its dual crystalline and superfluid nature by measuring the spectrum of its collective excitations. Moreover, by extending the experiments from one- to two-dimensional geometries, they have provided unambiguous evidence of its superfluidity by observing the formation of vortices when the supersolid is set into rotation. Many other properties of the supersolid phase have been investigated, including the order of the phase transition \cite{BiagioniPRX2022}, the effects of finite temperature on supersolidity \cite{Sanchez-BaenaNatCommun2022}, and measurements of the superfluid fraction of the gas \cite{BiagioniNature2024}, to cite just a few notable examples. Dipolar quantum gases therefore constitute without any doubt an ideal platform to explore supersolidity.

\subsection{Spin-orbit-coupled supersolids}
Dipolar supersolids are single-component systems in which supersolidity emerges due to the long-range dipole-dipole interaction, and where quantum fluctuations play a crucial role in stabilizing the system against collapse. Interestingly, bosonic mixtures with contact interactions in the pure mean-field regime can also give rise to a supersolid phase when subjected to spin-orbit coupling. In this section, we review the basic principles of spin-orbit coupling in quantum gases and explain how it leads to a supersolid phase in Bose-Einstein condensates. Next, we discuss how these systems are realized experimentally and how they allow one to probe both spontaneous density modulation and phase coherence. We then present recent experiments in which supersolidity of spin-orbit-coupled BECs is revealed through studies of the system's excitations. Finally, we conclude by outlining some of the interesting perspectives opened by the study of supersolidity in spin-orbit-coupled BECs.

\subsubsection{Theoretical description of spin-orbit-coupled Bose-Einstein condensates} \label{SOCtheory}

\begin{figure}[h]
\centering
\includegraphics[scale=1]{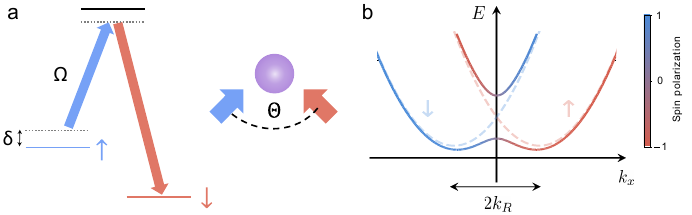}
\caption{\textbf{Experimental realization of spin-orbit coupling using a two-photon Raman process.} (a) Raman coupling scheme. Two internal states $\uparrow$ and $\downarrow$ are coupled \emph{via} a Raman transition with two-photon Rabi frequency $\Omega$ and detuning $\delta$. The laser beams have a relative angle of $\Theta$, leading to a momentum transfer associated with a spin flip along the $x$ direction. (b) Dispersion relation of the atom-photon dressed states for a coupling strength of $\Omega = 1.5 E_\mathrm{R}/\hbar$. Dashed lines show the single-particle dispersion relation of the bare states. Solid lines show the single-particle dispersion relation of the higher and lower dressed states. The color scale represents the spin composition (or spin polarization) of the dressed bands.} 
\label{fig:Raman} 
\end{figure}

\noindent\textbf{Spin-orbit coupling} \nopagebreak

\noindent\\
The key ingredient required to engineer a rotonized dispersion relation and obtain supersolidity in a bosonic quantum mixture is spin-orbit coupling. Spin-orbit coupling links the atoms' internal spin to their motional degrees of freedom. This coupling qualitatively alters their dispersion relation already at the single-particle level, giving rise to a momentum-dependent spin texture and, depending on the interactions between the uncoupled states, momentum-dependent interactions. Crucially, spin-orbit coupling also modifies the curvature of the bands, leads to multiple band minima and shifts their positions. In ultracold atoms, synthetic spin-orbit coupling is typically realized by coupling internal atomic states \emph{via} laser beams in such a way that a spin flip is accompanied by a momentum kick. Figure \ref{fig:Raman}a shows the most common scheme, where a set of Raman beams of wavelength $\lambda_\mathrm{R}$ couples two internal states $\uparrow$ and $\downarrow$ of energy splitting $\Delta\omega_{\mathrm{R}}$. The spatial configuration of the beams leads to a net momentum transfer $2 \hbar k_\mathrm{R} = 4 \pi \hbar \mathrm{sin}(\Theta /2)/\lambda_\mathrm{R}$ along one direction, which here we fix along the $x$ axis. In this expression, $\Theta$ is the relative angle of the two beams. The associated energy scale is the Raman recoil energy $E_\mathrm{R} = \hbar^2k_\mathrm{R}^2/(2m).$

This system is conveniently described in the spin-rotated frame with atomic basis $\{ | \uparrow, k+k_\mathrm{R}\rangle,| \downarrow, k-k_\mathrm{R}\rangle \}$, which is obtained by performing the unitary transformation $\mathcal{U}=\exp\left[i \left(\Delta \omega_\mathrm{R} t - 2k_{\mathrm{R}}x\right) {\sigma}_z/2 \right]$ on the initial atomic states. The single-particle Hamiltonian then takes the form 
\begin{equation}
\mathcal{H}_{\mathrm{SOC}}=\frac{\hbar^2}{2m} (\mathbf{k} + k_\mathrm{R} \sigma_z \mathbf{e_x} )^2 - \frac{\hbar \delta}{2} \sigma_z + \frac{\hbar \Omega}{2} \sigma_x, \label{H_SO} 
\end{equation}
where $\sigma_x$ and $\sigma_z$ are the Pauli matrices, and $\delta$ and $\Omega$ are the two-photon detuning and the Rabi frequency of the Raman-coupling process, respectively.  

Diagonalizing the Hamiltonian leads to two atom-photon dressed energy bands $\epsilon_\pm(\mathbf{k}) =\hbar^2(\mathbf{k}^2+k_\mathrm{R}^2)/(2m) \pm \hbar\tilde\Omega(k)/2,$ where $\tilde\Omega=\sqrt{\Omega^2 + \tilde\delta^2}$ and $\tilde\delta= \delta-2\hbar k_\mathrm{R}k/m$ are the generalized Rabi frequency and detuning. The energy bands correspond to a higher and a lower dressed state
\begin{align}
\ket{+, \mathbf{k}} &=  \mathcal{S}(k)\ket{\uparrow, \mathbf{k}} + \mathcal{C}(k)\ket{\downarrow, \mathbf{k}},\cr  
\ket{-, \mathbf{k}} &=  -\mathcal{C}(k)\ket{\uparrow, \mathbf{k}} + \mathcal{S}(k)\ket{\downarrow, \mathbf{k}},
\end{align}
where $\mathcal{C}(k) = \sqrt{(1 + \tilde{\delta}/\tilde{\Omega})/2}$, and $\mathcal{S}(k) = \sqrt{(1 - \tilde{\delta}/\tilde{\Omega})/2}$. Figure \ref{fig:Raman}b shows the dispersion relation and the momentum dependent spin composition of the dressed bands in the regime $\hbar\Omega < 4 E_\mathrm{R}$, where the lower dressed band features two minima. In the following we consider only this two-minima regime, as this is the situation relevant for supersolidity. Moreover, we assume that the BEC does not occupy the higher dressed band, which is separated from the lower one by an energy gap $\hbar \tilde{\Omega}$.

In a single particle picture, the position of the left ($\ell$) and right ($r$) minimum of the dispersion band $\epsilon_-$ are given by the recursive relation 
\begin{equation} \label{eq:minima}
    |k_i/k_\mathrm{R}| = \sqrt{1-[\hbar\Omega/(4E_\mathrm{R})]^2} 
\end{equation}
with $i=\ell, r$. Moreover, the spin composition of the dressed states at the minima is given by $ P = \tilde{\delta}(k_i)/\tilde{\Omega}.$ The Raman coupling therefore creates dressed particles that are superpositions of the bare states, and have a modified dispersion relation. As we will see below, it also modifies their effective interactions, which is the crucial ingredient to obtain a spin-orbit-coupled supersolid phase \cite{HigbiePRA2004, LinN2011, ChisholmScience2026}.\\

\noindent \textbf{Interactions and supersolidity} \\

\begin{figure}
\centering
\includegraphics[scale=1]{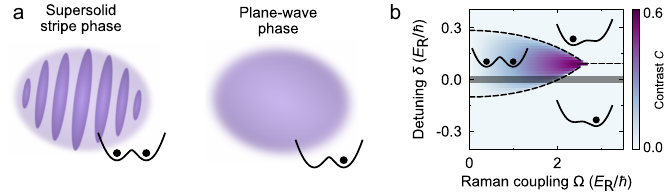}
\caption{\textbf{Supersolid-stripe phase and plane-wave phase of a spin-orbit-coupled BEC.} (a) The supersolid-stripe phase corresponds to the situation where both minima of the dispersion relation are occupied. It shows a periodic density modulation in the form of stripes. The density between the stripes does not go down to zero, meaning the contrast $C < 1$. The plane-wave phase corresponds to the situation where only one minimum is occupied and the density is unmodulated. (b) Phase diagram of a spin-orbit-coupled mixture of states $|F=1,m_F = 0\rangle$ and $|F=1,m_F = -1\rangle$ of a $^{41}$K at a magnetic field of $B=51.7$\,G, calculated \emph{via} the variational ansatz of the Trento group~\cite{LinPRL2009}. The phase boundaries (dashed lines) are obtained using the mixture model~\cite{ChisholmScience2026}, see main text. The plot shows the contrast of the density modulation depending on the Raman coupling strength and two-photon detuning. The contrast is finite in the supersolid stripe phase and drops to zero in the plane-wave phase. Figure adapted from \cite{ChisholmScience2026}.} 
\label{fig:supersolidvsplanewave} 
\end{figure}

\noindent In the two-minima regime of spin-orbit-coupled BECs, the effect of interparticle interactions leads to several possible ground states. They correspond to the occupation of only the left minimum, only the right minimum, or of both minima of the dispersion relation simultaneously. It is this last situation that gives rise to supersolidity. Due to the Raman dressing, the BECs sitting in the two minima are not fully orthogonal in their spin. There is therefore matter-wave interference between them, which produces a periodic (\emph{striped}) density modulation \cite{HigbiePRA2004, LinN2011, HoPRL2011}. In this situation, the translational symmetry is spontaneously broken: the spatial phase of the stripes depends on the relative phase of the two BECs, which is different in each experimental realization. The contrast of the interference pattern increases with the spin overlap of the two states, which increases with the Raman coupling strength $\Omega$. At the same time, the system remains superfluid and spontaneously breaks U(1) gauge symmetry. In contrast, when a single minimum is occupied, the system is in the so-called plane-wave phase, which does not have any density modulation and therefore no supersolidity. Both scenarios are summarized in Fig. \ref{fig:supersolidvsplanewave}a. \\

\noindent \textbf{Phase diagram} \\

\noindent A powerful way to understand the phase diagram of a spin-orbit-coupled BEC as a function of the spin-orbit coupling strength $\Omega$, and including also the effect of a detuning $\delta$, is the variational approach developed by the Trento group \cite{LiPRL2012}. The idea is to minimize the energy functional of the spin-orbit-coupled system 
\begin{equation}
E=\int\mathrm d\mathbf{r}\left[\Psi^{\dagger}\mathcal{H}_{\mathrm{SOC}}\Psi+\frac{g_{\uparrow\uparrow}}{2}\left|\psi_{\uparrow}\right|^2+\frac{g_{\downarrow\downarrow}}{2}\left|\psi_{\downarrow}\right|^2+g_{\uparrow\downarrow}\left|\psi_{\uparrow}\right|^2\left|\psi_{\downarrow}\right|^2\right]
\end{equation}
using an ansatz for the two-component spinor wavefunction which is a superposition of two plane waves
\begin{equation}
\Psi=\begin{pmatrix}\psi_\uparrow \\
\psi_\downarrow
\end{pmatrix}
=
\sqrt{n}\left[C_1
\begin{pmatrix}
-\sin\theta_1 \\
\cos\theta_1
\end{pmatrix}
e^{i \kappa_1 x}+C_2
\begin{pmatrix}
-\cos\theta_2 \\
\sin\theta_2
\end{pmatrix}
e^{-i \kappa_2 x}\right]\label{eq:variationalPsi}.
\end{equation}
This ansatz has six variational parameters: the angles $\theta_1$ and $\theta_2$ that quantify the spin polarization of the two momentum components of the wavefunction, the wavevectors $\kappa_1$ and $\kappa_2$ of the plane waves, which are related to the polarization as $\kappa_i=k_R\cos{2\theta_i}$, and the complex weights $C_1$ and $C_2$, which are normalized as $|C_1|^2+|C_2|^2=1$. Equation \eqref{eq:variationalPsi} corresponds to the exact ground state of the system in the absence of interactions, with $\kappa_i$ coinciding with the single-particle values $k_i$ of Eq. \eqref{eq:minima}. It is therefore natural to expect that it can accurately describe the system as long as interactions remain weak, as indeed demonstrated in \cite{LinPRL2009}. The variational approach yields the phase diagram of Fig.  \ref{fig:supersolidvsplanewave}b, which is plotted for the specific interaction parameters of $^{41}$K atoms. It shows a dome of finite modulation contrast corresponding to the supersolid-stripe phase, surrounded by plane-wave regions where only one minimum of the dispersion is occupied. The minimum that is favored depends on the value of the two-photon detuning and of the intraspin interactions. The tip of the supersolid dome corresponds to the situation where the two plane-wave phases have the same energy and become equally probable. In the figure, it is not centered at $\delta=0$ because for $^{41}$K the two intraspin interactions are different, $g_{\uparrow\uparrow}\neq g_{\downarrow\downarrow}$.\\

\begin{figure}
\centering
\includegraphics[scale=1]{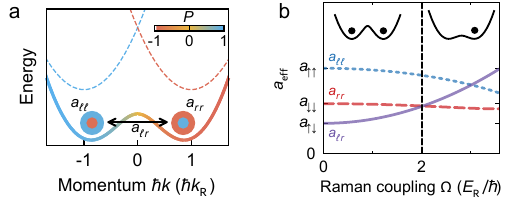}
\caption{\textbf{Effective interactions in the mixture model.}  (a) Dressed BECs in the mixture model. The Raman-dressed BECs occupy the minima of the dispersion and interact with the effective scattering lengths $a_{\ell\ell}$, $a_{rr}$ and $a_{\ell r}$. These effective interactions arise from the spin composition of the dressed states, illustrated schematically as a large blue (red) circle with a smaller red (blue) circle inside for the left (right) minimum. (b) Phase diagram for detuning $\delta = 0$. The plot shows how the effective interactions of the dressed BECs evolve as the Raman coupling increases. The bare-state interactions  $a_{\uparrow\uparrow}$, $a_{\downarrow\downarrow}$ and $a_{\uparrow\downarrow}$ correspond to the same values as in Fig. \ref{fig:supersolidvsplanewave}. Once the inter-well scattering length $a_{\ell r}$ exceeds any of the intra-well values $a_{\ell\ell}$ or $a_{rr}$, the system exits the supersolid stripe phase and transitions into the plane-wave phase. Figure adapted from \cite{ChisholmScience2026}.} 
\label{fig:mixture-model} 
\end{figure}

\noindent \textbf{Effective mixture description} \\

\noindent In the two-minimum regime, spin-orbit-coupled BECs can also be described as an effective bosonic mixture \cite{HigbiePRA2004, LinN2011, ChisholmScience2026}. Within the mixture approach, the system is treated as a mixture of dressed BECs that occupy the two minima left ($\ell$) and right ($r$) of the dispersion relation and have renormalized properties due to the Raman coupling. Specifically, they acquire a modified effective mass $m^*$ given by the curvature of the dispersion, and effective scattering lengths $a_{\ell\ell}$, $a_{rr}$ and $a_{\ell r}$ given by the polarization of the dressed states at the dispersion minima, see Fig. \ref{fig:mixture-model}a. 

As a function of the bare interaction parameters of the system and the Raman coupling parameters, which enter through the momenta of the minima $k_j$, they take the form
\begin{align}\label{Seq_aij}
a_{jj} = &a_{\uparrow\uparrow}\frac {(1-k_j/k_{\mathrm{R}})^2}4 + a_{\downarrow\downarrow}\frac {(1+k_j/k_{\mathrm{R}})^2}4 +
a_{\uparrow\downarrow} \frac{(1-k_j^2/k_{\mathrm{R}}^2)}2 ,\cr
a_{\ell r} = 
&a_{\downarrow\downarrow}\frac {(1+k_{\ell}/k_{\mathrm{R}})(1+k_r/k_{\mathrm{R}})}2 \cr + 
&a_{\uparrow\uparrow}\frac {(1-k_{\ell}/k_{\mathrm{R}})(1-k_r/k_{\mathrm{R}})}2 \cr + 
&a_{\uparrow\downarrow}\frac{1 - k_{\ell} k_r/k_{\mathrm{R}}^2 + \sqrt{1-k_{\ell}^2/k_{\mathrm{R}}^2}\sqrt{1-k_r^2/k_{\mathrm{R}}^2}}{2},
\end{align}
with $j=\ell, r$. This result can be formally obtained by developing an effective field theory of the system truncated around the minima of the dispersion \cite{ChisholmScience2026}.

From these scattering lengths, the phase diagram of the system displayed in Fig. \ref{fig:supersolidvsplanewave}b can be understood as that of an effective mixture. The system is in the supersolid-stripe phase when it is energetically favorable to occupy the two minima $\ell$ and $r$, which leads to interference and density modulation. This occurs when the two dressed BECs are miscible $a_{\ell r}^2 <a_{\ell\ell}a_{rr}$ and $a_{\ell r}<a_{\ell \ell}, a_{rr}$.  Otherwise, it is preferable to occupy a single minimum, the system polarizes and transitions to the plane-wave phase, and the density modulation disappears. Because the effective scattering lengths depend on the Raman-coupling parameters, which control the polarization of the dressed states at the minima, the transition from miscibility to polarization depends on the values of $\Omega$ and $\delta$. The situation for $\delta=0$ and the parameters of $^{41}$K is depicted in Fig. \ref{fig:mixture-model}b. Interestingly, the mixture model not only gives an intuitive picture of the mechanisms at play, but also provides analytical expressions for the boundaries of the phase diagram \cite{ChisholmScience2026}.

\subsubsection{Experiments with spin-orbit-coupled supersolids}

\noindent\textbf{First experiments} \\

\begin{figure}[t]
\centering
\includegraphics[scale=1]{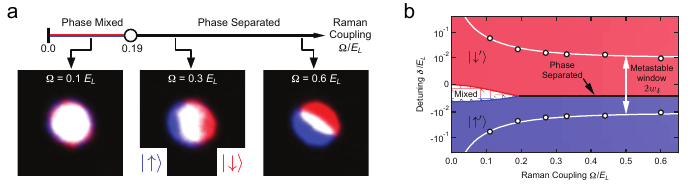}
\caption{\textbf{Phase separation in spin-orbit-coupled $^{87}$Rb BECs.} (a) Miscible-immiscible transition of the dressed spins. The three example images show absorption images of the bare spin states. For imaging, the dressed states were first mapped onto the bare spin states before the atoms were released from the trap. The images therefore reflect the in situ density distribution of the dressed components. As the Raman coupling strength increases, the two components undergo a transition from a miscible to a phase-separated configuration and spatially separate. The residual spatial overlap observed in the phase-separated regime arises from the harmonic confinement of the trapping potential (b) Phase diagram of the system. It shows the mixed and phase separated states, corresponding to the stripe and plane-wave phases, as well as the metastable window in which the experimentally prepared spin polarization does not relax to the ground state value. Figure adapted from \cite{LinN2011}.} 
\label{fig:phase-separation} 
\end{figure}

\noindent The investigation of spin-orbit-coupled BECs was pioneered by the group of I. Spielman at JQI. The first experiments were performed using $^{87}$Rb atoms \cite{LinN2011}. They revealed two dressed spin states with distinct quasi-momentum at coupling strength $\Omega\hbar < 4E_\mathrm{R}$, which merged into a single dressed component with only one quasi-momentum upon increase of the Raman coupling strength. Although the bare spin states are miscible, a transition from a miscible to an immiscible mixture was observed at $\hbar\Omega = 0.19 E_\mathrm{R}$, see Fig. \ref{fig:phase-separation}a. These experimental measurements reveal how the effective interactions between the dressed BECs are modified by the spin-orbit coupling, as discussed in the previous section. Indeed, as soon as $a_\mathrm{\ell r}^2<a_\mathrm{\ell \ell}a_\mathrm{rr}$ it becomes energetically favorable to spatially separate the dressed states. Another important experimental observation was the existence of large metastable windows where the spin populations did not correspond to those of the ground state, and could not adjust to the correct values within the duration of the experiment, see Fig. \ref{fig:phase-separation}b. Finally, although the existence of the stripe phase was already predicted in this work, it could not be observed. The main challenges were the vanishingly small contrast of the stripes $C<5\%$ and, more importantly, the very small area of the phase diagram that the phase occupies. In practice, achieving the level of Raman detuning control required to observe supersolidity in 
$^{87}$Rb demands exceptionally stable magnetic fields, which is extremely challenging to realize experimentally.\\

\noindent\textbf{Probing broken translational symmetry and phase coherence}\\

\noindent To increase the area that the supersolid stripe phase occupies in the phase diagram, the group of W. Ketterle at MIT followed an alternative approach \cite{LiNature2017}. Instead of coupling two internal states, they used as pseudospins the orbital states of an asymmetric double well potential, which was created with an optical superlattice. Due to the small spatial overlap of such orbital states, their interspin interaction $a_{\uparrow\downarrow}$ is small. This leads to a substantial improvement of the phase diagram, because  the miscible regime reaches larger coupling strengths and therefore significant stripe contrasts. Moreover, in this scheme the orbital states are magnetically insensitive, improving the control over the detuning and therefore the stability of the stripe phase. However, the experiment was limited to small coupling strengths. Because the energy splitting of the two pseudospins is comparable to other energy scales of the system, the Raman coupling leads to Floquet heating. In practice, this limited the stripe contrast to $C \leq 8\%$. 

\begin{figure}[t]
\centering
\includegraphics[scale=1]{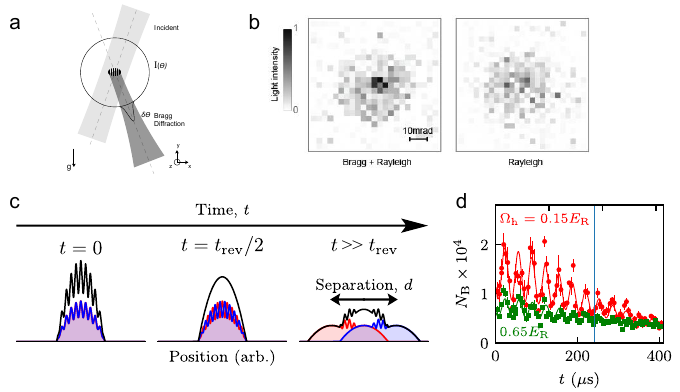}
\caption{\textbf{Density modulation and phase coherence in a spin-orbit-coupled BEC.} (a) Optical Bragg diffraction scheme. A beam incident at angle $\Theta$ is diffracted by the supersolid density modulation, which acts as a grating. (b) First observation of supersolidity in a spin-orbit-coupled condensate. The left image shows the Bragg scattering signal from the supersolid, while the right image shows the Rayleigh scattering signal present even without spin-orbit coupling.
(c) Talbot interferometry sequence. The cloud with periodic density modulation is released from the trap. If the two components shift by half a modulation period ($t_{\mathrm{rev}}/2$), they are out of phase, producing a flat density profile. At $t_{\mathrm{rev}}$, the components are in phase again, interfering to recreate the modulated density pattern. At longer times, their spatial overlap diminishes. (d) Phase coherence. The plot shows the visibility of the modulation pattern as a function of time, measured \emph{via} optical Bragg scattering. Periodic revivals demonstrate long-range phase coherence. For stronger Raman coupling (green data points), the signal is reduced due to decreased miscibility. Figures adapted from \cite{LiPhD2020,LiNature2017,PutraPRL2020}.} 
\label{fig:first-SOCSS} 
\end{figure}

Experimentally observing the density modulation of the stripe phase is challenging. The modulation not only has a very small contrast but also a small spatial period on the order of $1/k_{\mathrm{R}}$, corresponding to just a few hundred nanometers. To detect it, the MIT group developed an optical Bragg scattering method. The approach relies on observing the diffraction of light caused by the density modulation, which acts as a grating, see Fig. \ref{fig:first-SOCSS}a. Experimentally, they could indeed observe a sharp diffraction peak on top of a diffuse Rayleigh background. The resulting signal, shown in Fig. \ref{fig:first-SOCSS}b, provided the first demonstration of the spontaneous breaking of translational symmetry in the stripe phase of spin-orbit-coupled BECs.

Supersolidity requires not only a spontaneous density modulation but also phase coherence between the stripes. The group of I. Spielman exploited a Talbot interferometer to demonstrate that spin-orbit-coupled BECs are indeed phase coherent along the direction of the stripes \cite{PutraPRL2020}. The experiment used a Raman-coupled $^{87}$Rb BEC, in which the two dispersion minima were occupied either in the ground state or in metastable stripe configurations. The density modulation was observed using the optical Bragg scattering technique of \cite{LinN2011}, and to enhance the Bragg signal, the spin-orbit coupling strength was rapidly increased just prior to detection. Phase coherence was then probed \emph{via} matter-wave Talbot interferometry, adapting a scheme originally developed for optical lattice systems \cite{MiyakePRL2011, SantraNatCommun2017} to the spin-orbit-coupled case. The measurements, shown in Fig. \ref{fig:first-SOCSS}d, confirm that the stripe phase exhibits long-range phase coherence.

Although the MIT and JQI experiments demonstrate the broken translational symmetry of the stripe phase, unlike in the dipolar supersolid experiments reported in Sec. \ref{sec:dipolarsupersolids} the density modulation could not be imaged directly. As discussed above, the two main challenges were the vanishingly small contrast of the stripes and their submicron spacing. At ICFO, we recently overcame these limitations \cite{ChisholmScience2026} and demonstrated direct imaging of the stripes. To increase the contrast of the stripes, we used $^{41}$K atoms, where the relative strength of the inter- and intraspin interactions can be adjusted using Feshbach resonances \cite{TanziPR2018}. 
By tuning the magnetic field to a point of low interspin interactions, the stripe phase is less sensitive to the two-photon detuning and extends to higher Raman coupling, with a stripe contrast $C>50\%$. The corresponding phase diagram is the one shown in Fig. \ref{fig:supersolidvsplanewave}b of Sec. \ref{SOCtheory}. To overcome the optical resolution of standard imaging systems, we  magnified the in situ density distribution using a matter-wave magnification technique, adapting previously demonstrated schemes \cite{ShvarchuckPRL2002,MurthyPRA2014,AsteriaN2021}, to spin-orbit-coupled systems. 

\begin{figure}[t]
\centering
\includegraphics[scale=1]{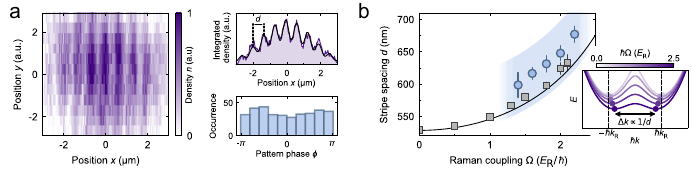}
\caption{\textbf{In situ observation of a spin-orbit-coupled supersolid.} (a) In situ images of a spin-orbit-coupled BEC in the supersolid stripe phase. The image is taken using a matter-wave magnification technique and shows a periodic density modulation. The integrated profile is fitted with a modulated Gaussian. The histogram shows the phase of the density modulation over 350 realizations. (b) Stripe spacing $d$ as a function of the Raman coupling strength $\Omega$. It is measured from both in situ images (blue circles) and time-of-flight ones (grey squares). The solid black line is the theory prediction corresponding to $d=2\pi/\Delta k$, where $\Delta k$ is the momentum difference of the two minima. The inset illustrates the dispersion relation and the position of its two minima for different coupling strengths.  Figure adapted from \cite{ChisholmScience2026}.} 
\label{fig:insitustripe} 
\end{figure}

Figure \ref{fig:insitustripe}a shows the resulting in situ picture of the supersolid stripe phase. The integrated density profile is used to characterize the cloud in terms of stripe spacing and cloud size. Figure \ref{fig:insitustripe}b shows the experimentally measured stripe spacing $d$ for different strengths of the Raman coupling $\Omega$. The spacing grows with coupling strength, which is a result of the changing dispersion relation. Indeed, as depicted in the inset, with increasing coupling strength the two minima of the dressed band reduce their momentum difference $\Delta k$. If the dressed states are located at the minima of the dispersion, $d\propto 1/\Delta k$ and thus the stripe spacing grows with $\Omega$. \\

\noindent\textbf{Collective excitations}\\

\begin{figure}
\centering
\includegraphics[scale=1]{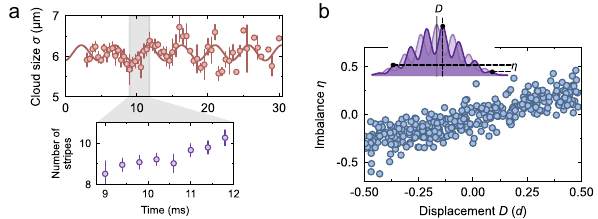}
\caption{\textbf{Dynamics of a spin-orbit-coupled supersolid.} (a) Breathing mode of the stripe phase. The cloud size is measured from in situ images and plotted versus time. The solid line shows a sinusoidal fit to the data. The inset below shows the number of stripes in the system. As the cloud size grows, more stripes are dynamically created in the cloud, since the stripe spacing is independent from the cloud size. (b) Statistical analysis of the displacement of the stripe pattern versus imbalance of the edge stripes. When the spatial phase of the density modulation moves, it is compensated by a superfluid counterflow that redistributes the atoms between the stripes. The linear relation between the displacement and the imbalance provides evidence of the low-energy crystal Goldstone mode in a harmonically trapped system. Figure adapted from \cite{ChisholmScience2026}.} 
\label{fig:superfluid-flow} 
\end{figure}

\noindent The investigation of collective excitations turned out to be crucial to characterize the supersolid phase in dipolar systems, as discussed in Sec. \ref{sec:dipolarsupersolids}. In spin-orbit-coupled BECs, theoretical works have predicted a rich excitation spectrum with crystal and superfluid modes deriving from the spin and density degrees of freedom of the cloud \cite{GeierPRL2021, MartoneSciPost2021, GeierPRL2023}. At ICFO, we have leveraged our ability to directly image the density distribution of spin-orbit-coupled supersolids to investigate the supersolid dynamics and probe both its superfluid and crystalline properties \cite{ChisholmScience2026}. Figure \ref{fig:superfluid-flow}a shows the collective breathing mode of the cloud extracted from the cloud size. The measured frequency is in accordance with superfluid hydrodynamics, which predicts a breathing mode  frequency $\omega_\mathrm{B} = \sqrt{5/2}\omega_\mathrm{D}$, where $\omega_\mathrm{D}$ is the dipole mode frequency of the system. Besides the breathing of the cloud, the experiment revealed a dynamic change in the amount of stripes visible. Since the cloud size is not coupled to the stripe spacing, additional stripes are created and extinguished as the cloud expands and contracts. This reveals a superfluid flow across the system and shows that particles can move between stripes. It also demonstrates an important difference to dipolar systems, where the system size and stripe spacing are not independent and the crystal and superfluid modes mix.

The experiment also allows to observe signatures of the crystal Goldstone mode. Similarly to the case of dipolar supersolids discussed in Sec.\ref{sec:dipolar_excitations}, the harmonic trap breaks the degeneracy of different spatial positions of the modulation pattern and the crystal Goldstone mode acquires a finite energy. Then, a sliding of the crystal is accompanied by a superfluid counter flow that redistributes the density between the stripes to counteract the phase shift and keep the center of mass unchanged. Figure \ref{fig:superfluid-flow}b shows the relation between displacement of the center stripe and imbalance of the population of the side stripes. The resulting linear dependence is characteristic of the low-energy Goldstone mode. Interestingly, in the experiment the energy penalty for displacing the stripes $\sim0.2$ Hz is small enough for the mode to be excited by the system preparation without a targeted excitation scheme. This value is more than one order of magnitude smaller than in typical dipolar supersolid experiments \cite{GuoNature2021, BiagioniNature2024}.

\begin{figure}
\centering
\includegraphics{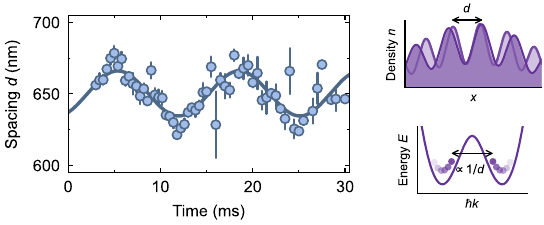}
\caption{\textbf{Observation of a crystal compression mode in a spin-orbit-coupled supersolid.} The plot shows the stripe spacing oscillating depending on holding time.
The solid line is a sinusoidal fit to the data. The illustrations show the stripe compression mode in real space, as an oscillation of the stripe spacing, and in momentum space as an out-of phase oscillation of the two dressed BECs in the dispersion minima. It is thus equivalent to a spin-dipole mode of the effective mixture. Figure adapted from \cite{ChisholmScience2026}.} 
\label{fig:compression_mode} 
\end{figure}

Various studies on the excitation spectrum of supersolid-like systems have revealed significant differences between the platforms. One of the most striking aspects is the nature of the crystal structure. 
While dipolar supersolids can host phonons, single-mode cavity systems have a rigid crystal with a modulation spacing fixed by the light wavelength. In these systems, only multimode cavities have been able to restore phonon modes \cite{GuoNature2021}. Since studies on spin-orbit-coupled supersolids have long been limited to optical Bragg scattering, the dynamics of the crystal structure in the stripe phase was considered inaccessible, and, despite recent theoretical clarifications \cite{GeierPRL2023}, the question whether spin-orbit-coupled systems could host phonons had remained experimentally unresolved. With the first in situ observation and access to the supersolid dynamics, however, the collective modes can now be studied. At ICFO, we have observed a crystal compression mode that is a dynamic oscillation of the stripe spacing.  Figure \ref{fig:compression_mode} shows the experimental data as extracted from the in situ images. 
A fast ramp of the Raman coupling excites an out-of-phase dipole oscillation of the two dressed states, see inset of Fig. \ref{fig:compression_mode}. As the dressed states oscillate in the two minima, they change their momentum difference in time and thus the spacing of the interference pattern oscillates. For this reason, this mode can also be interpreted as a spin-dipole mode of the dressed states. The observation of the crystal compression mode demonstrates that spin-orbit-coupled supersolids have a compressible crystal structure and can host phonon modes like a real solid. In this aspect they are similar to dipolar supersolids.

\begin{figure}
\centering
\includegraphics[scale=1]{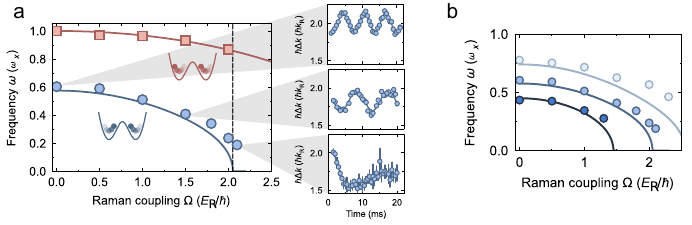}
\caption{\textbf{Locating the supersolid phase transition \emph{via} mode softening.}(a) Frequencies of the dipole mode (red squares) and stripe compression mode (blue circles) for a spin-orbit-coupled supersolid. The compression-mode frequency softens at the phase transition from the supersolid stripe to the plane-wave phase. The solid lines are predictions of the mixture model without any fitting parameters. The insets show exemplary oscillations measured at different spin-orbit coupling strengths. (b) Tunability of the phase transition. The compression mode frequency softening is measured for three different values of the magnetic field, which correspond to different values of the inter- and intraspin scattering lengths. This modification of the interactions shifts the location of the phase transition point. Figure adapted from \cite{ChisholmScience2026}.} 
\label{fig:mode_softening} 
\end{figure}

The frequency of the stripe compression mode has been predicted to soften when approaching the phase transition from stripe to plane-wave phase, completely vanishing in the plane-wave phase \cite{GeierPRL2021}. This frequency reduction upon increasing Raman coupling has been observed experimentally at ICFO, see Fig. \ref{fig:mode_softening}a. In contrast, the in-phase dipole mode of the dressed components, which correspond to the dipole mode of the atoms in the trapping potential, is not affected by the phase transition. It however shows a small reduction of its frequency with increasing coupling. Within a mixture picture, it can be traced to the change of effective mass of the dressed states. As experimentally shown by the USTC group, its frequency goes to zero at the plane-wave to single minimum transition \cite{ZhangPRL2012}.

The frequency softening of the stripe compression mode can be used to locate the supersolid phase transition, see dashed line in Fig. \ref{fig:mode_softening}a. Moreover, the critical coupling strength that marks the transition point is determined by the relation of intraspin to interspin interactions. This effect can be seen in Fig. \ref{fig:mode_softening}b, where the value of the bare state interactions have been tuned using a Feshbach resonance. Reducing the interspin interactions $a_{\uparrow\downarrow}$ displaces the transition point towards higher coupling strength.

\subsection{Conclusions and outlook}
While the concept of supersolidity originally emerged in the context of quantum solids, its experimental existence remained controversial for many years. Ultracold atomic gases have since provided a powerful platform to study supersolid-like phenomena. Thanks to their microscopic controllability and the precise tuning of particle interactions, supersolidity has become an experimentally accessible and highly tunable many-body phase. In this review we have focused on dipolar gases and spin-orbit-coupled mixtures. These two platforms have realized supersolidity as a ground-state phase with a spontaneous density modulation that can host phonons reminiscent of real solids. Taken together, these two systems demonstrate that supersolid phases can arise from distinct microscopic mechanisms. Their comparison provides insight into which features are generic consequences of simultaneously broken gauge and translational symmetries and which are artifacts of each specific system.

In dipolar gases, long-range interactions lead to the self-organized formation of density-modulated states. These systems have provided an in-depth investigation of dipolar supersolids over the last years, characterizing their ground state and excitation spectrum, and providing definite evidence of their superfluid nature through the observation of quantized vortices. 

Spin-orbit-coupled systems realize a conceptually different route, in which the structure of the single-particle dispersion and interactions give rise to roton softening and the emergence of a spontaneous density modulation. Despite being one of the first platforms to report hallmarks of supersolidity, a quantitative characterization of these states was challenged for a long time by the extreme fragility of the phase, small modulation contrast and short spacing of the modulation pattern. These challenges have however been recently overcome, and observations of the density modulation and phase coherence of spin-orbit-coupled supersolids have recently been complemented with studies of their collective excitations spectrum \cite{ChisholmScience2026}. 

Collective excitations have indeed been investigated across many quantum-gas platforms that display supersolid-like features \cite{LéonardS2017, TanziN2019, NatalePRL2019, GuoN2019, GuoNature2021, TanziScience2021, NorciaPRL2022, LiebsterNatPhys2025, ChisholmScience2026}. In particular, compressional modes and the existence of a crystal phonon branch have been revealed in dipolar and spin-orbit-coupled supersolids. Due to the available excitation schemes and limited system sizes, the experiments could however only perform a global excitation. The observation of local excitations, similar to those studied in driven superfluids \cite{LiebsterNatPhys2025}, would be a natural extension of these works. Another exciting research direction is the study of the crystal Higgs mode, which corresponds to a periodic oscillation of the crystal's contrast and has so far only been observed in cavity systems \cite{LéonardS2017}. It would be very interesting to explore it in dipolar and spin-orbit-coupled systems, and recent theoretical works have been preparing such studies. In dipolar supersolids, the Higgs mode hybridizes with higher excited modes in the trapping geometries investigated so far \cite{HertkornPRL2019}, leading to strong damping and hindering its experimental observation. Indeed, the competition between trap geometry, long-range and contact interactions intertwines the crystalline character of a dipolar supersolid with its density envelope. One proposed solution is a ring-shaped trap geometry \cite{HertkornPRR2024, MukherjeePRL2025}. In spin-orbit-coupled systems, spin modes and density modes are mostly decoupled. The crystal Higgs mode, which is of spin character, is thus expected to be more easily accessible \cite{ChisholmScience2026}, and further theoretical work is currently underway to confirm this scenario. 

While in the first dipolar supersolids the spontaneous crystallization of the system occurred only along one dimension, more recent experiments have realized two-dimensional supersolid crystals \cite{NorciaN2021, BlandPRL2022}. Spin-orbit-coupled supersolids remain for the moment restricted to the one-dimensional scenario. In Barcelona, we have already devised a practical scheme to engineer spin-orbit-coupled condensatess with two dimensional crystals by adding a third atomic state and another pair of Raman-coupling beams. However, these theoretical ideas still need to be realized experimentally. Interestingly, such experiments would reveal a fundamental difference between dipolar and spin-orbit-coupled systems. For isotropic harmonic traps, dipolar supersolids break spontaneously both the translational and the rotational symmetry, meaning that the orientation of the two-dimensional crystal takes different values in each experimental realization. In contrast, in the spin-orbit-coupled case the crystallization direction will be fixed by the geometric layout of the Raman-coupling beams, realizing a phase that might be closer to a superfluid smectic than to a true supersolid. 

Finally, two recent highlights in the field of dipolar supersolids have been the observation of quantized vortices in the system, which provides a smoking gun for superfluidity \cite{CasottiN2024}, and the measurement of a reduced superfluid fraction along the direction of the supersolid crystal \cite{BiagioniNature2024}. Developing practical schemes to prove the superfluid character of spin-orbit-coupled supersolids beyond the measurement of the breathing mode frequency, and to determine their superfluid fraction, which in this case could be measured much closer to the thermodynamic limit, remain therefore important goals for the near future. 

\section{Conclusion}

In these lecture notes, we have reviewed two closely related research directions in the field of bosonic quantum mixtures with tunable interatomic interactions. First, we discussed the stabilization of attractive Bose-Bose mixtures against collapse through the repulsive effect of quantum fluctuations, which gives rise to ultradilute quantum liquids and self-bound quantum droplets. Second, we presented the stripe phase: a supersolid phase of matter that emerges in bosonic mixtures with repulsive interactions under Raman-induced spin-orbit coupling. In both cases, we outlined the theoretical framework, summarized key experimental results, and highlighted open questions for future investigation.
 
Dipolar quantum gases share strong conceptual connections with both the beyond-mean-field stabilization and the supersolid behavior observed in spin-orbit-coupled bosonic mixtures. Because these systems display many analogous phenomena and have developed largely in parallel, we have also discussed some aspects of dipolar gases. In these systems, the interplay between contact and dipole-dipole interactions can stabilize quantum droplets even at the single-component level. The droplets can organize into ordered arrays and, under suitable conditions, exchange particles to establish global phase coherence, forming a supersolid phase. Throughout the lecture notes, we have emphasized both the similarities and distinctions between dipolar and mixture systems.

A key difference lies in the role of quantum fluctuations. In dipolar systems, quantum fluctuations are essential: dipolar supersolids cannot be described without beyond-mean-field effects, which stabilize the droplet arrays. By contrast, spin-orbit-coupled supersolids are conceptually simpler and can be captured within a mean-field framework. This fundamental distinction drives much of the differing phenomenology between the two platforms.

Isolating the effect of quantum fluctuations independently of the dipolar system is therefore of great interest. In spin-orbit-coupled BECs, the mean-field energy can be tuned to near cancellation by adjusting the interactions between the dressed condensates to an attractive regime ($a_{\ell r} < 0$). In such a regime, quantum liquid droplets are expected to form, exhibiting supersolid properties such as spontaneous density modulation \cite{SachdevaPRA2020, Sanchez-BaenaPRA2020}. Realizing this scenario would provide independent control over beyond-mean-field effects, offering a unique window into the role of quantum fluctuations in supersolidity.

Finally, the creation of dipolar mixtures represents another exciting frontier at the intersection of dipolar gases and bosonic mixtures. The additional species enriches the system dynamics and enables novel structures, including alternating domain supersolids \cite{BlandPRA2022} and self-bound droplet crystals \cite{ArazoPRR2023}. Moreover, the second component can catalyze supersolidity in an otherwise unmodulated dipolar BEC by modifying the effective dipolar interactions \cite{ScheiermannPRA2023}. Exploring these systems promises new opportunities to investigate tunable supersolid phases and the interplay between competing interactions, quantum fluctuations, and multi-component physics.

\begin{acknowledgement}
We acknowledge all members of the Quantum Gases Experimental (QGE) group at ICFO that have contributed over the years to the development of this research line. In particular, for the quantum liquid droplet experiments C. R. Cabrera, P. Cheiney, J. Sanz and L. Tanzi, and for the supersolid experiments C. S. Chisholm, V. Makhalov, R. Ramos, and R. Vatr\'e. We would also like to highlight the contribution of J. Cabedo and A. Celi, from Universitat Aut\`onoma de Barcelona, to our understanding of spin-orbit-coupled supersolids through the development of the mixture model. Finally, we acknowledge many insightful discussions over the years on quantum droplet physics and supersolidity with G. Astrakharchik, J. Boronat, T. Bourdel, T. Donner, F. Ferlaino, I. Ferrier-Barbut, W. Ketterle, T. Langen, J. L\'eonard, G. Martone, G. Modugno, D. Petrov, T. Pfau, A. Recati, I. Spielman, S. Stringari, and L. Santos.

Our work on these topics is currently supported by the European Union (ERC CoG-101003295 SuperComp), the Spanish Ministry of Science and Innovation MCIU/AEI/10.13039/501100011033 (projects MAPS PID2023-149988NB-C22, and Severo Ochoa CEX2024-001490-S), Generalitat de Catalunya (CERCA program), Fundaci\'o Cellex and Fundaci\'o Mir-Puig. Moreover, S.H. acknowledges support from the European Union (Marie Sklodowska Curie–101149245 Epiquant).
\end{acknowledgement}

\ethics{Competing Interests}{The authors have no conflicts of interest to declare that are relevant to the content of this chapter.}

\eject
\bibliographystyle{apsrev4-2}
%\bibliography{references_ch_Tarruell.bib}

%

\end{document}